%% file: main.tex
\documentclass[twoside]{article}

\usepackage[accepted]{aistats2024} 

\usepackage[round]{natbib}

\usepackage[blackhypersetup]{shortex} 
\usepackage{xcolor}
\usepackage{bbm}
\usepackage{xr}

\DeclareMathOperator{\proj}{proj}
\parsefontstylestrings{{c}}{SW}

\begin{document}

%

%

\twocolumn[

\aistatstitle{Coreset Markov chain Monte Carlo}

\aistatsauthor{Naitong Chen \And Trevor Campbell}

\aistatsaddress{Department of Statistics \protect\\ University of British Columbia \protect\\ \texttt{naitong.chen@stat.ubc.ca}\And 
                Department of Statistics \protect\\ University of British Columbia \protect\\ \texttt{trevor@stat.ubc.ca}} ]

\input{abstract.tex}

\input{introduction.tex}
\input{background.tex}

\input{coresetmcmc.tex}

\input{experiments.tex}

\input{conclusion.tex}

\subsubsection*{Acknowledgements}
We acknowledge the support of an NSERC Discovery Grant (RGPIN-2019-03962), and
the use of the ARC Sockeye computing platform from the
University of British Columbia.

\bibliographystyle{unsrtnat}
\bibliography{main.bib}

\section*{Checklist}

 \begin{enumerate}

 \item For all models and algorithms presented, check if you include:
 \begin{enumerate}
   \item A clear description of the mathematical setting, assumptions, 
   algorithm, and/or model. \textbf{Yes, this may be found in \cref{sec:coresetmcmc}.} 
   \item An analysis of the properties and complexity (time, space, sample size) 
   of any algorithm. \textbf{Yes, this may be found in \cref{sec:coresetmcmc}, with more detailed in 
   \cref{sec:proofs,sec:appendix_gaussian_example}.}
   \item (Optional) Anonymized source code, with specification of all dependencies, 
   including external libraries. \textbf{Yes, this is included in the zip file in the supplementary material.}
 \end{enumerate}

 \item For any theoretical claim, check if you include:
 \begin{enumerate}
   \item Statements of the full set of assumptions of 
   all theoretical results. \textbf{Yes, this may be found in \cref{sec:coresetmcmc}.}
   \item Complete proofs of all theoretical results. \textbf{Yes, this may be found in 
   \cref{sec:proofs,sec:appendix_gaussian_example}.}
   \item Clear explanations of any assumptions. \textbf{Yes, the assumptions are listed in \cref{sec:coresetmcmc}, 
   with more detailed explanations in \cref{sec:proofs,sec:appendix_gaussian_example}.}
 \end{enumerate}

 \item For all figures and tables that present empirical results, check if you include:
 \begin{enumerate}
   \item The code, data, and instructions needed to reproduce the main experimental 
   results (either in the supplemental material or as a URL). \textbf{Yes, 
   this is included in the zip file in the supplementary material, and a URL is also provided in \cref{sec:experiments}.}
   \item All the training details (e.g., data splits, hyperparameters, how they were 
   chosen). \textbf{Yes, this may be found in \cref{sec:expt}.}
         \item A clear definition of the specific measure or statistics and error 
         bars (e.g., with respect to the random seed after running experiments 
         multiple times). \textbf{Yes, the error metrics are defined in \cref{sec:experiments}, 
         and the error bars are explained in the caption of each figure.}
         \item A description of the computing infrastructure used. (e.g., type of 
         GPUs, internal cluster, or cloud provider). \textbf{Yes, this may be found in \cref{sec:experiments}.}
 \end{enumerate}

 \item If you are using existing assets (e.g., code, data, models) or 
 curating/releasing new assets, check if you include:
 \begin{enumerate}
   \item Citations of the creator if your work uses existing assets. 
   \textbf{Yes, these are included as footnotes in \cref{sec:expt}.}
   \item The license information of the assets, if applicable. \textbf{Not Applicable. We did not use any licensed assets.}
   \item New assets either in the supplemental material or as a URL, if applicable. \textbf{Yes,  
   this is included in the zip file in the supplementary material, and a URL is also provided in \cref{sec:experiments}.}
   \item Information about consent from data providers/curators. \textbf{Not Applicable. 
   We did not use any data that required consent from providers/curators.}
   \item Discussion of sensible content if applicable, e.g., personally 
   identifiable information or offensive content. \textbf{Not Applicable. None of such information is included in the paper.}
 \end{enumerate}

 \item If you used crowdsourcing or conducted research with human subjects, 
 check if you include:
 \begin{enumerate}
   \item The full text of instructions given to participants and screenshots. 
   \textbf{Not Applicable. No human subjects are involved in this paper.}
   \item Descriptions of potential participant risks, with links to Institutional 
   Review Board (IRB) approvals if applicable. \textbf{Not Applicable.}
   \item The estimated hourly wage paid to participants and the total amount 
   spent on participant compensation. \textbf{Not Applicable.}
 \end{enumerate}

 \end{enumerate}

\appendix
\onecolumn

\input{proofs.tex}

\input{gaussian.tex}

\input{experimental_details.tex}

\input{additional_results.tex}


\end{document}

%% file: abstract.tex
\begin{abstract}
A Bayesian coreset is a small, weighted subset of data that replaces the
full dataset during inference in order to reduce computational cost.
However, state of the art methods for tuning coreset weights are
expensive, require nontrivial user input, and impose constraints on the model.
In this work, we propose a new method---\emph{Coreset MCMC}---that 
simulates a Markov chain targeting the coreset posterior, while simultaneously updating
the coreset weights using those same draws. Coreset MCMC 
is simple to implement and tune, and can be used with any existing MCMC kernel. 
We analyze Coreset MCMC in a representative setting to obtain key insights about the convergence
behaviour of the method. Empirical results demonstrate that Coreset MCMC
provides higher quality posterior approximations and reduced computational cost
compared with other coreset construction methods. Further, compared with other 
general subsampling MCMC methods, we find that Coreset MCMC has a higher sampling 
efficiency with competitively accurate posterior approximations.
\end{abstract}

%% file: introduction.tex
\section{\uppercase{Introduction}}
\label{sec:introduction}
Bayesian inference provides a flexible and principled framework for parameter estimation and uncertainty 
quantification when working with complex statistical models. Markov chain Monte Carlo (MCMC) (\citealp{robert1999monte}; 
\citealp{robert2011short}; \citealp[Ch.~11,12]{gelman2013bayesian}) is the standard methodology 
for conducting Bayesian inference, and involves simulating a Markov chain whose stationary distribution is the
Bayesian posterior. However, in the large-scale data setting, standard MCMC methods become costly, as simulating each Markov transition 
involves iterating over the entire data set; and typically, many steps are needed to obtain reasonable estimates
of posterior expectations. 

To reduce the computational cost incurred by large datasets, \emph{subsampling  
MCMC} methods repeatedly subsample the data and simulate the next state of the chain
based on only that subsample, instead of the full dataset 
\citep{banterle2019accelerating,quiroz2018speeding,maclaurin2014firefly,korattikara2014austerity,bardenet2014towards,
welling2011bayesian,chen2014stochastic}. Some of these methods target the exact posterior
distribution asymptotically, but require a full pass over the data per accepted step \citep{banterle2019accelerating}, 
require model-specific design of log-likelihood surrogates \citep{maclaurin2014firefly},
or have progressively slower convergence \citep{welling2011bayesian}.
Other methods approximate various aspects of the transition with 
a subsample---e.g., the accept-reject decision \citep{korattikara2014austerity,bardenet2014towards},
or dynamics that generate a proposal \citep{chen2014stochastic,baker2019sgmcmc}---but the performance
of such methods is dependent on the existence of good control variates to keep (gradient) log-likelihood estimate variance
low \citep{quiroz2018speeding}. Control variate design is model-specific in general, while automated methods 
impose limitations on the model (e.g., Taylor expansion-based control variates require continuous latent variables). For more in-depth 
reviews of subsampling MCMC methods, see \cite{bardenet2017markov,quiroz2018subsampling,nemeth2021stochastic}.

Another approach to handling large-scale Bayesian inference problems are \emph{Bayesian coresets} \citep{huggins2016coresets,campbell2019automated,campbell2018bayesian,campbell2019sparse,
manousakas2020pseudocoresets,naik2022fast,chen2022bayesian}, which consist of a fixed, weighted subsample of the data that replaces 
the full dataset during inference. The intuition is that there is typically
redundancy present in large data, and so it should be possible to capture the information needed for inference 
in a single small subsample. Indeed, recent work has shown that a coreset of size $O(\log N)$ suffices to 
provide a near-exact approximation of a posterior for $N$ data points in a wide class of models (\citealp[Thm.~4.1,4.2]{naik2022fast}; \citealp{chen2022bayesian}).
Coresets do not require bespoke kernel or control-variate design; once a 
coreset is constructed, any generic MCMC kernel can be applied. Coresets also typically preserve important structure (e.g., unidentifiability, heavy tails)
as they are built using the original model's log-likelihood terms. 

Constructing a good coreset efficiently
remains a challenge. Early methods based on importance sampling 
\citep{huggins2016coresets} are simple and computationally efficient, but generally produce unreliable approximations in practice.
Methods based on sparse regression with a finite-dimensional log-likelihood projection 
\citep{campbell2019automated,campbell2018bayesian,zhang2021revisiting} are sensitive to the 
projection and require significant expert user tuning effort.
Sequential greedy KL minimization approaches \citep{campbell2019sparse} are able to produce high 
quality coresets, but involve a slow and difficult-to-tune inner-loop weight optimization with outer-loop point selection.
Two recent methods---Sparse Hamiltonian Flows (SHF) and Quasi-Newton Coresets (QNC) \citep{naik2022fast,chen2022bayesian}---instead 
quickly select the points in the coreset via uniform subsampling, and
then run a single joint weight optimization. SHF avoids inner-outer loop procedures entirely, but is limited 
to differentiable log-posterior densities and does not come with a convergence guarantee on the weights. QNC
applies more broadly and has a guaranteed weight convergence, but only if certain expectations can be evaluated exactly;
in practice, one requires inner-loop MCMC. Both methods tend to require significant user expertise and effort when tuning.

This work introduces \emph{Coreset MCMC}, a new construction method that is simpler and faster to implement and tune
compared with previous coreset methods. Coreset MCMC can be thought of as a meta-algorithm that wraps
an existing MCMC kernel: the method iterates between (1) taking a step with the kernel targeting the coreset posterior, and (2) adapting the coreset
weights using the current draw, and is related to both adaptive stochastic approximation algorithms \citep[see][pp.~31-33]{benveniste2012adaptive}
and adaptive MCMC \citep[see][]{andrieu2008tutorial}. We show that when the optimal coreset is exact, Coreset MCMC will produce
a coreset that converges to the exact posterior, and hence is an asymptotically exact method. We also analyze Coreset MCMC 
using a representative model to obtain key insights in tuning the method. This paper concludes with experiments demonstrating that coreset 
MCMC provides higher quality posterior approximations than other coreset methods, and improved sampling efficiency 
compared with subsampling MCMC methods.

%% file: background.tex
\section{\uppercase{Background}}
\label{sec:background}
We are given a dataset $(X_n)_{n=1}^N$ of $N$ 
observations, a log-likelihood $\ell_n(\theta) \defas \log p(X_n \given \theta)$ for observation $n$ given 
$\theta \in \Theta$, and a prior density $\pi_0(\theta)$. The goal is to sample from the Bayesian posterior with density
\[
  \pi(\theta) \defas \frac{1}{Z} \exp\left( \sum_{n=1}^N \ell_n(\theta) \right) \pi_0(\theta),
\]
where $Z$ is the normalizing constant. The Bayesian coresets approach to reducing this cost involves replacing the 
sum of log-likelihoods for the full dataset with a small, weighted subsample (without loss of generality, we assume
these are the first $M$ points $1, \dots, M$):
\[
  \pi_w(\theta) \defas \frac{1}{Z(w)} \exp\left( \sum_{m=1}^M w_m \ell_m(\theta) \right) \pi_0(\theta),
\]
where $w \in \reals^M, \, w\geq 0$ is a vector of nonnegative weights. 
In this work, we follow the setting in \citet{naik2022fast} and \citet{chen2022bayesian}, 
where the coreset points are uniformly subsampled.
If we can construct a set of weights $w$ 
such that $M\ll N$ and $\pi_w \approx \pi$, then we can generate draws using MCMC targeting $\pi_w$ 
as a surrogate for draws from $\pi$ at a low per-iteration cost.

Most recent coreset construction methods formulate the task as a variational inference
problem, following \citet{campbell2019sparse}:
\[
  w^\star = \argmin_{w\in\reals^M} \kl{\pi_w}{\pi} \quad \text{s.t.} \quad w \in\fcW.\label{eq:coresetopt}
\]
In this work, we assume that the feasible region $\fcW \subseteq \reals^M$ is convex and closed. We also assume
that for all $w \in \fcW$, $Z(w) < \infty$.
In practice, $\fcW$ is usually the set of nonnegative weights $\reals_+^M$, but often also includes other constraints, e.g., $\sum_m w_m = N$.
This variational problem is also slightly unusual, in the sense that we cannot estimate the KL divergence \emph{even up to a constant}
because of the unknown normalization $Z(w)$ that depends on the coreset weights $w$. However, we can write
the $M$-dimensional KL gradient as an expectation under $\pi_w$ that does not explicitly involve the normalization:
\[
  &\nabla_w \kl{\pi_w}{\pi} \\
  = &\cov_{\pi_w}\lt( \bbmat \ell_1(\theta) \\ \vdots \\ \ell_M(\theta)\ebmat, \sum_m w_m\ell_m(\theta) - \sum_n \ell_n(\theta) \rt). \label{eq:grad}
\] 
If we had access to \iid draws from $\pi_w$, we could obtain an unbiased estimate of the gradient in \cref{eq:grad} for stochastic gradient descent 
\citep{robbins1951stochastic,Bottou2004} by estimating the covariance using these \iid draws,
and estimating the full-data sum $\sum_n \ell_n(\theta)$ with a subsample.
Note that crucially, this estimate can be computed using
only black-box evaluations of the log-likelihood, which makes this approach applicable to a wide range of models (e.g., with discrete variables).
However, \iid draws from $\pi_w$ are usually not available, and one must resort to MCMC.
This introduces an inner loop MCMC estimation step, which is slow and may need constant tuning 
as the coreset weights $w$ evolve \citep{campbell2019sparse,naik2022fast}.
In the next section, we draw inspiration from adaptive MCMC and adaptive stochastic approximation algorithms to develop
a coreset construction method that more naturally interleaves MCMC and gradient updates.

%% file: coresetmcmc.tex
\section{\uppercase{Coreset MCMC}}
\label{sec:coresetmcmc}

\subsection{Setup}
Suppose at iteration $t\in\nats$, for a fixed set of coreset weights $w_t\in\fcW$,
we had access to $K\geq 2$ \iid draws $\theta_t = \lt(\theta_{t1}, \dots, \theta_{tK}\rt) \in \Theta^K$ from $\pi_{w_t}$, 
and a subsample $\fcS_t \subseteq [N]$ of $S$ data points
drawn uniformly without replacement from the full dataset.
Then an unbiased estimate of the gradient in \cref{eq:grad} is given by
\[
&g(w_t, \theta_t, \fcS_t) = \label{eq:gradest}\\
&\frac{1}{K-1}\sum_{k=1}^K \!\!\bbmat \bar\ell_1(\theta_{tk})\\ \vdots \\ \bar\ell_M(\theta_{tk})\ebmat\!\!\lt(\sum_m w_{tm}\bar\ell_m(\theta_{tk}) \!-\! \frac{N}{S}\sum_{s\in \fcS_t}\bar\ell_{s}(\theta_{tk}) \rt),
\]
where $\bar\ell_n(\theta_{tk})$ are the centered log-likelihoods 
\[
\bar\ell_n(\theta_{tk}) = \ell_n(\theta_{tk}) - \frac{1}{K}\sum_{j=1}^K \ell_n(\theta_{tj}).
\]
Since the estimate $g(w_t,\theta_t,\fcS_t)$ is unbiased, i.e.,  $\ex g(w_t,\theta_t, \fcS_t) = \grad\kl{\pi_{w_t}}{\pi}$,
it can be used in a stochastic optimization scheme:\footnote{In \cref{eq:sgd,eq:coresetmcmcopt} we show standard
projected stochastic gradient descent, but other methods such as ADAM \citep{kingma2014adam} and AdaGrad \citep{duchi2011adaptive} are also possible. We 
use ADAM in our experiments.}
\[
w_{t+1} = \proj_{\fcW}\lt( w_t - \gamma_t g(w_t, \theta_t, \fcS_t)\rt), \label{eq:sgd}
\]
where $\gamma_t > 0$ is a monotone decreasing learning rate sequence, and
$\proj_{\fcW}$ projects the result onto the feasible set $\fcW$.
However, in realistic scenarios, we do not have access to independent draws from $\pi_w$;
we are forced to take approximate draws using Markov chains.
Our algorithm in \cref{sec:alg} is based on using the stochastic optimization scheme in \cref{eq:sgd}
but with $K$ Markov chains instead of independent sequences.

\subsection{Algorithm}\label{sec:alg}
\balg[t]
\caption{One iteration of Coreset MCMC} \label{alg:coresetmcmc}
\balgc
\Require $w_t$, $\theta_t$, $\gamma_t$, $\kappa_w$, $S$
\LineComment Subsample the data
\State $\fcS_t \gets \distUnif\lt(S, [N]\rt)$ (without replacement)
\LineComment Compute gradient estimate
\State $g_t \gets $ \cref{eq:gradest}
\LineComment Stochastic gradient step
\State $w_{t+1} \gets w_t - \gamma_t g_t$
\LineComment Step each Markov chain
\For{$k=1, \dots, K$}
\State $\theta_{(t+1)k} \dist \kappa_{w_{t+1}}(\cdot \given \theta_{tk})$
\EndFor
\State\Return $w_{t+1}$, $\theta_{t+1}$
\ealgc
\ealg

\textbf{Iteration:} Let $\kappa_w$ be a family of Markov kernels parametrized by coreset weights $w$ such that
for each set of weights $w$, $\kappa_w$ has invariant distribution $\pi_w$.
Then one step of Coreset MCMC, shown in \cref{alg:coresetmcmc}, 
involves the following updates in sequence:
\[
w_{t+1} &\gets \proj_{\fcW}\lt(w_t - \gamma_t g(w_t, \theta_t, \fcS_t)\rt) \label{eq:coresetmcmcopt}\\
\theta_{(t+1)k} &\dist \kappa_{w_{t+1}}\lt(\cdot \given \theta_{tk}\rt), \quad k=1,\dots,K,
\]
where $\fcS_t$ is a subset of $S$ indices from $\{1, \dots, N\}$ drawn uniformly without replacement.
Note that $K \geq 2$ Markov chains are required for the gradient estimate 
in \cref{eq:gradest}; each step in these $K$ chains is independent given the current weights, and can be run in parallel.

\textbf{Initialization:} We initialize the coreset weights uniformly
to have sum $N$:
\[
w_{0m} &= \frac{N}{M} & m&=1,\dots,M,
\]
and initialize the Markov chain states $\theta_{0k}$, $k=1,\dots,K$ arbitrarily,
although it may help in practice to allow some amount of burn-in before beginning
to adapt the coreset weights.

\textbf{Choice of kernel:} Coreset MCMC is agnostic as to the choice of kernels $\kappa_w$; 
for example, $\kappa_w$ could be a basic random-walk Metropolis--Hastings scheme (\citealp[Ch.~7]{robert1999monte}), slice sampler \citep{neal2003slice},
or Hamiltonian Monte Carlo-based method \citep{neal2011mcmc,duane1987hybrid} designed to target $\pi_w$. 
However, because the weights $w_t$---and hence the coreset posterior $\pi_{w_t}$---will change as iterations proceed, 
one should prefer kernels that automatically adapt to the local geometry of the coreset posterior. For example, a hit-and-run slice sampler 
with doubling (\citealp[Fig.~4]{neal2003slice}; \citealp{belisle1993hit}) is a reasonable choice as it will automatically select reasonable step sizes as $w$ changes.
Although adaptive MCMC kernels \citep[see][]{andrieu2008tutorial} may seem attractive here, they are usually designed for a single target---not a moving target
like $\pi_{w_t}$---and tuning such methods jointly with Coreset MCMC may be difficult.

\textbf{Complexity:} If the kernel $\kappa_w$ can be applied in $O(M)$ time, 
and each log-likelihood term $\ell_n(\theta)$ takes $O(1)$ time to compute,
each step of Coreset MCMC takes $O((M+S)K)$ time on a single processor.
If $K$ processors are available, this can be reduced to $O((M+S)\log K)$
by running the Markov chains in parallel and using distributed reduction in \cref{eq:gradest}.

\textbf{Subsampling:} We draw $S_t$ without replacement, as this 
reduces subsampling variance by $\frac{N-S}{N-1} \leq 1$ versus sampling
with replacement. However, when $S \ll N$, the difference between these two is small.
In practice, one can perform random order scans over the data.

\subsection{Convergence Analysis} \label{sec:convanalysis}
We now analyze the convergence behaviour of Coreset MCMC. Proofs may be found in \cref{sec:proofs}.
So far we have assumed that $\gamma_t > 0$ is a (not necessarily strictly) monotone decreasing sequence,
$\fcW$ is closed and convex, and that $\forall w\in\fcW$, $Z(w) < \infty$; all of these assumptions
are required to use \cref{alg:coresetmcmc}.
In this subsection, we impose a set of additional assumptions 
that are not required by \cref{alg:coresetmcmc} but simplify the analysis.

Note that the coreset optimization problem \cref{eq:coresetopt} is nonconvex;
but despite this fact, we are able to show that Coreset MCMC 
produces an optimal set of weights in settings where
$M$ is large enough such that there is an exact coreset.
We formalize this assumption in \cref{assump:unique}.
\bassump[Exact coreset]\label{assump:unique}
There exists a unique $w^\star \in \reals^M$, $c^\star\in\reals$
such that $w^\star \in \fcW$ and
\[
\sum_{n=1}^N \ell_n(\cdot) = \sum_{m=1}^M w^\star_m\ell_m(\cdot) + c^\star \qquad \pi_0-a.e.v.
\]
\eassump

We now begin with the simple case where $S=N$, i.e., where there is no data subsampling in the stochastic gradient estimate \cref{eq:gradest}.
Define $G_t\in\reals^{M\times M}$ to be 
\[
G_t &= \frac{1}{K-1}\sum_{k=1}^K
\bbmat \bar\ell_1(\theta_{tk})\\ \vdots \\ \bar\ell_M(\theta_{tk})\ebmat
\bbmat \bar\ell_1(\theta_{tk})\\ \vdots \\ \bar\ell_M(\theta_{tk})\ebmat^\top,
\]
and note that the KL gradient estimate in \cref{eq:gradest} can be written
\[
g(w_t, \theta_t, [N]) &= G_t\lt(w_t - w^\star\rt).
\]
Hence all of the stochasticity in the gradient estimate in \cref{eq:gradest}
comes from $G_t$, which estimates $C_{w_t} = \cov_{\pi_{w_t}}\bbmat \ell_1(\theta) & \dots & \ell_M(\theta)\ebmat$. 
As is usual in the analysis of stochastic optimization, we require
two high-level conditions for convergence: the gradient estimate must (1) provide 
progress per iteration on average,
and (2) not be so noisy that the algorithm
makes unrecoverable mistakes. We formulate these
in \cref{assump:mixing,assump:bounded} in terms of the moments of $G_t$ 
after a single Markov chain step.
\bassump[Markov gradient mixing]\label{assump:mixing}
There exists $\lambda > 0$ such that 
\[
\forall w_t &\in \fcW, \theta_{t-1} \in \Theta^K & \ex \lt[ G_t \given w_t, \theta_{t-1} \rt] &\succeq \lambda I.
\]
\eassump
\bassump[Markov gradient noise boundedness]\label{assump:bounded}
There exists a $\overline{\lambda} < \infty$ such that
\[
\forall w_t \in \fcW, \,\,\theta_{t-1} \in \Theta^K \,\,\,\, 
\ex \!\lt[G_t^\top G_t\!\given\! w_t, \theta_{t-1}\rt] \preceq \overline{\lambda} I.
\]
\eassump
\cref{assump:mixing} can be interpreted in two parts. First, we ask that
the Markov chains mix quickly so that the covariance estimate $G_t$ is similar 
to the exact covariance $C_{w_t}$ in the KL gradient formula \cref{eq:grad}. 
Second, we require the exact covariance $C_{w_t}$ to have a
positive minimum eigenvalue to guarantee that \cref{alg:coresetmcmc} progresses towards the optimum.
\cref{assump:bounded} can be interpreted as placing a bound on the variance of $G_t$
at each iteration. 
Both \cref{assump:mixing,assump:bounded} are 
meant to be representative, akin to the uniformly bounded gradient noise
assumptions common in the optimization literature \citep[e.g.][]{rakhlin2011making,hazan2014beyond};
they are too strong to hold precisely in realistic models with unbounded parameter spaces $\Theta$.
However, note that \cref{assump:bounded} holds
if the $\ell_n$ are continuous and the parameter space $\Theta$ is compact;
and \cref{assump:mixing} holds if additionally $\fcW$ is compact and the minimum eigenvalue of $\ex\lt[ G_t \given w_t, \theta_{t-1}\rt]$ 
is continuous and strictly positive for $w_t\in\fcW$, $\theta_{t-1}\in\Theta^K$.

\cref{thm:converge_full} shows that Coreset MCMC produces weights that converge
to the optimum $w^\star$ in expectation (and hence in probability) when the full data are
used for gradient estimation, i.e., when $S=N$ in \cref{eq:gradest}.  In
particular, a constant learning rate $\gamma_t=\gamma$ in
\cref{thm:converge_full} yields linear convergence  despite the use of
stochastic gradient estimates. This is because the gradient variance shrinks as
the iterates approach the optimum due to the multiplicative $(w_t - w^\star)$
factor in the gradient estimate. 
The link between variance reduction and linear convergence in stochastic gradient methods
is well-known in the optimization literature; see \cite{gower2020variance} 
for a more in-depth discussion. 
\bthm\label{thm:converge_full}
Suppose \cref{assump:unique,assump:mixing,assump:bounded} hold and $S=N$.
There exists $\gamma > 0$ such that if $\sup_t\gamma_t \leq \gamma$, 
\[
\ex\|w_t - w^\star\|^2 \leq e^{-\lambda\sum_{\tau=0}^{t-1}\gamma_\tau}\|w_0 - w^\star\|^2.
\]
In particular, if $\sum_{t=0}^{\infty}\gamma_t = \infty$, then $\ex\|w_t - w^\star\|^2 \to 0$ as $t \to \infty$.
\ethm
When gradients are estimated using subsampling, i.e.,
when $S < N$ in \cref{eq:gradest}, there is an additional
noise term that must be controlled in the analysis. Define
\[
\Delta_{tkn} &= \bar\ell_n(\theta_{tk}) - \frac{1}{N}\sum_{n'=1}^N\bar\ell_{n'}(\theta_{tk})\\
V_t &= \frac{1}{MN}
	\sum_{m=1}^M\sum_{n=1}^N\lt(\frac{1}{K-1}\sum_{k=1}^K\bar\ell_m(\theta_{tk})\Delta_{tkn}\rt)^2. \label{eq:subnoise}\\
\]
The quantity $V_t \geq 0$ captures the variance of subsampling noise.
\cref{assump:noisebounded} uniformly bounds the influence of this noise after a single Markov chain step.
Again, although meant just to be representative, \cref{assump:noisebounded} holds at least when
the log-likelihood functions are continuous and the parameter space $\Theta$
is compact.
\bassump[Subsampling noise boundedness]\label{assump:noisebounded}
There exists a $V < \infty$ such that
\[
\forall w_t \in \fcW, \,\,\theta_{t-1} \in \Theta^K \qquad 
\ex \lt[V_t\given w_t, \theta_{t-1}\rt] \leq V.
\]
\eassump
Given the addition of \cref{assump:noisebounded}, \cref{thm:converge_subsample} provides a convergence
guarantee for Coreset MCMC with data subsampling.
Note that convergence is now sublinear due to subsampling noise.

\begin{figure*}
\begin{subfigure}{0.24\textwidth}
\includegraphics[width=\columnwidth]{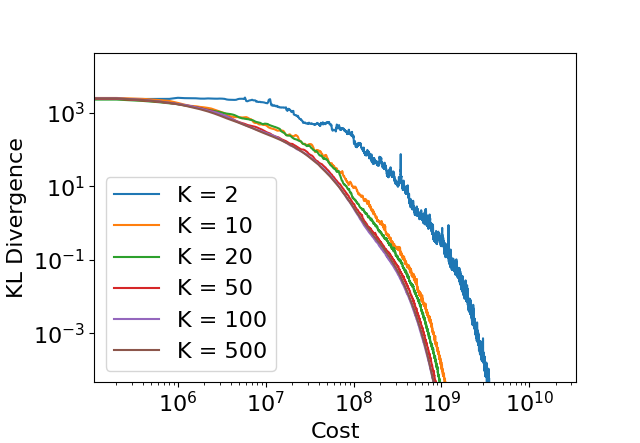}
\caption{Varying $K$, full data}\label{fig:heuristicKfull}
\end{subfigure}
\begin{subfigure}{0.24\textwidth}
\includegraphics[width=\columnwidth]{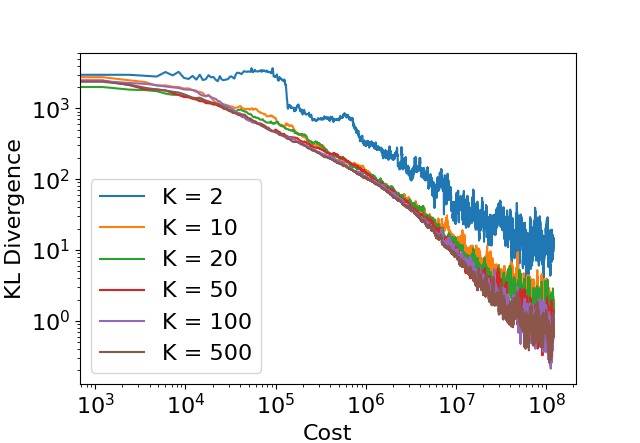}
\caption{Varying $K$, subsampling}\label{fig:heuristicKsubsample}
\end{subfigure}
\begin{subfigure}{0.24\textwidth}
\includegraphics[width=\columnwidth]{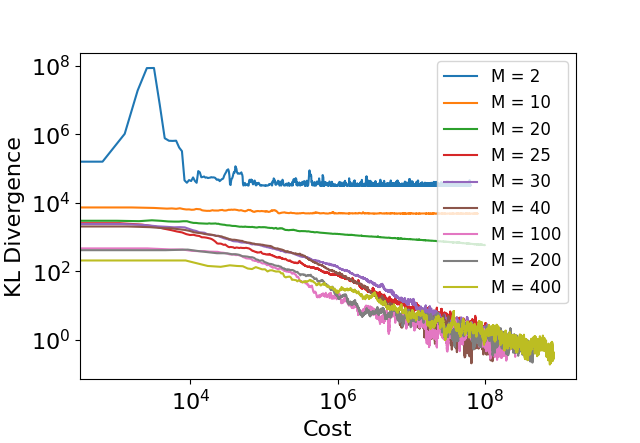}
\caption{Varying $M$}\label{fig:heuristicM}
\end{subfigure}
\begin{subfigure}{0.24\textwidth}
\includegraphics[width=\columnwidth]{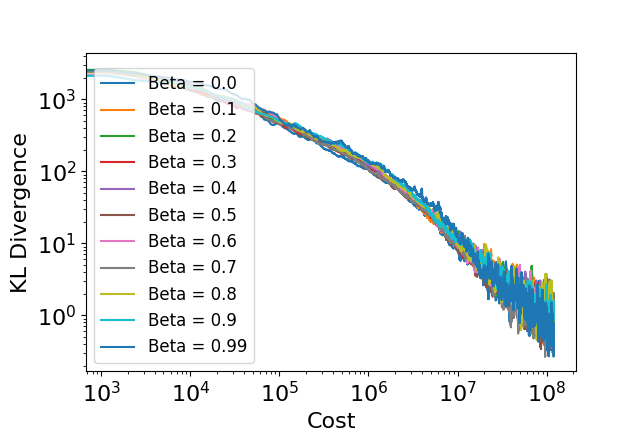}
\caption{Varying $\beta$}\label{fig:heuristicbeta}
\end{subfigure}
\caption{Influence of gradient estimation scheme (with or without subsampling), number of Markov chains, Markov chain
mixing rate, and coreset size on the performance of Coreset MCMC in the
Gaussian location model.
}\label{fig:heuristic}
\end{figure*}

\bthm\label{thm:converge_subsample}
Suppose \cref{assump:unique,assump:mixing,assump:bounded,assump:noisebounded} hold.
There exists $\gamma >0$ such that if $\sup_t\gamma_t \leq \gamma$, then
\[
\ex \|w_{t} - w^\star\|^2 
&\leq e^{-\lambda \sum_{\tau=0}^{t-1}\gamma_\tau}\|w_0-w^\star\|^2 \\
&+ \frac{V(N-S)N^2M}{(N-1)S}\sum_{\tau=0}^{t-1}\gamma_\tau^2 e^{-\lambda \sum_{u=\tau+1}^{t-1}\gamma_u}.
\]
In particular, if $\sum_{t=0}^{\infty}\gamma_t = \infty$ and $\gamma_t \to 0$ as $t\to\infty$,
then $\ex\|w_t - w^\star\|^2 \to 0$ as $t \to \infty$.
\ethm

\subsection{Intuition from a Representative Model}\label{sec:gaussianlocation}
In this subsection we build intuition on the behaviour of \cref{alg:coresetmcmc}
using a Gaussian location model in $\reals^d$,
with prior $\theta \dist \distNorm(0,I)$, data $X_n \distiid \distNorm(\theta, I)$,
$N=10{,}000$, and $d=20$.
We use the feasible region $\fcW = \{w\in\reals^M : w\geq 0, \sum_m w_m = N\}$.
The coreset posterior is $\pi_w = \distNorm(\mu_w, \sigma^2 I)$ with $\mu_w = \sigma^2 \sum_{m=1}^M w_m X_m$ and
$\sigma^2 = \frac{1}{1+N}$.
We use a family of Markov kernels $\kappa_w$ inspired by Langevin dynamics,
\[
\theta_{(t+1)k} &\dist \distNorm\lt(\sqrt{\beta}(\theta_{tk}-\mu_{w_t}) + \mu_{w_t}, (1-\beta) I\rt),
\]
where $\beta \in [0,1]$ controls how quickly the Markov chains mix; $\beta = 0$
provides independent draws, $\beta = 1$ yields no mixing. 
We use a default setting of $\beta = 0.8$, $S = 30$, $M = 30$, $K = 20$ for each experiment,
and vary one parameter while holding the others fixed. 
We use a learning rate of the form $\gamma_t = \gamma(t+1)^{\alpha-1}$, with 
$\gamma = \frac{N}{10 M}$ and---following
\cref{thm:converge_full,thm:converge_subsample}---$\alpha = 1$ for full-data gradient estimates
and $\alpha = 0.5$ for subsampled estimates.
We plot KL divergence computed using \cref{eq:gaussiankl} versus ``cost'' equal to $((M+S)\log K)t$, per the earlier complexity analysis.
A heuristic analysis of the expected
KL divergence when $\beta \approx 0$, $N\gg M\gg 1$, and $\gamma \ll M$ (see
\cref{sec:appendix_gaussian_example}) yields the formula
\[
&\ex \kl{\pi_{w_t}}{\pi} \label{eq:heuristic}
\lessapprox \\
&e^{-\frac{2\gamma\lt(t^\alpha - 1\rt)}{\alpha }}\frac{N}{2M} 
+ \frac{\gamma(N-S)d(K+d)(1+\log t^{1-\alpha})}{4S(K-1)t^{1-\alpha}}.
\]

\textbf{Effect of parallelization ($\bm{K}$):} Inspecting the formula in \cref{eq:heuristic}, 
we expect increasing the
number of chains should not influence the expected KL significantly for full-data gradients,
and when subsampling should provide a meaningful reduction until $K \approx d$.
\cref{fig:heuristicKfull,fig:heuristicKsubsample} confirm this intuition. 
We recommend setting $K$ as large as possible given available hardware in practice, as it may
improve approximation quality without a noticeable time penalty.  

\textbf{Effect of subsampling ($\bm{S}$):} \cref{fig:heuristicKfull,fig:heuristicKsubsample}
also demonstrate the influence of subsampled gradient estimates, and confirm the results in
\cref{eq:heuristic} and \cref{thm:converge_full,thm:converge_subsample}. In particular, the KL divergence
appears to converge geometrically when using the full data gradient estimates, and polynomially when
using subsampled estimates. However, adjusting for iteration cost, subsampling can provide improved approximation 
quality, especially in earlier iterations. A reasonable default setting of $S$ in practice is such that the time spent simulating
the Markov chains and computing the gradient update is comparable.

\textbf{Effect of coreset size ($\bm{M}$):} \cref{fig:heuristicM} shows that Coreset MCMC still converges
when the optimal coreset is not exact, but that there is persisting error. Once the coreset gets large enough
such that the optimal coreset is exact, further increasing the size does not yield major improvements.
In this problem, $M\approx d = 20$ suffices to get an exact coreset, and
beyond that the effect of increasing $M$ further diminishes.

\textbf{Effect of mixing rate ($\bm{\beta}$):} \cref{fig:heuristicbeta} shows that, at least in this example, the mixing
rate of the Markov chain has little effect on the performance of Coreset MCMC. In practice, the mixing rate 
of any kernel $\kappa_w$ could in principle be adjusted by increasing the number of steps between gradient updates.
Given the results here, we recommend using a single application of an automatically adapting kernel, e.g., a slice sampler with doubling
\citep{neal2003slice}.

%% file: experiments.tex
\section{\uppercase{Experiments}}\label{sec:experiments}
In this section, we compare \texttt{CoresetMCMC} (without subsampling, i.e., $S=N$) and \texttt{CoresetMCMC-S} (with subsampling, i.e., $S < N$) 
against previous coreset construction methods, as well 
as various subsampling MCMC methods.
The coreset construction methods that we compare to are uniform subsampling 
(\texttt{Unif})---which assigns a weight of $N/M$ to $M$ uniformly drawn points---as well as 
\texttt{QNC} \citep{naik2022fast} and \texttt{SHF} \citep{chen2022bayesian}. 
Note that we do not compare to Sparse VI \citep{campbell2019sparse} due to its high computational cost; 
see \citet{naik2022fast} for a comparison between Sparse VI and \texttt{QNC}.
The subsampling MCMC methods we compare to are Austerity MH 
(\texttt{Austerity}) \citep{korattikara2014austerity}, confidence MH (\texttt{Confidence}) 
\citep{bardenet2014towards}, and stochastic gradient Langevin dynamics (\texttt{SGLD-CV})
and stochastic gradient Hamiltonian Monte Carlo (\texttt{SGHMC-CV}) each with 
a control variate set to the log-likelihood gradient near the posterior density mode \citep{nemeth2021stochastic}.

We consider four Bayesian regression models: a (non-conjugate) linear, a logistic, a Poisson, 
and a sparse regression model.
The models for the first three real data experiments contain only continuous variables, and that for the last
synthetic experiment contains both continuous and discrete variables; see \cref{sec:expt} for details. 
We use Stan \citep{carpenter2017stan} to obtain full data inference results for real data experiments, 
and use the Gibbs sampler developed by \cite{george1993variable} for the synthetic experiment, due to
its inclusion of discrete variables.

For the three real data experiments (with only continuous variables),
we measure the posterior approximation quality of all methods 
using the two-moment KL, defined as
  $\kl{\distNorm( \hat{\mu}, \hat{\Sigma} )}{\distNorm( \mu, \Sigma ) }$,
where $\hat{\mu}, \hat{\Sigma}$ are the mean and covariance estimated using draws from each method,
and $\mu, \Sigma$ are the same estimated for the full data posterior.
This metric combines posterior mean and covariance error into a single number.
We measure efficiency using wall-clock training time as well as
the minimum marginal effective sample size (ESS) per second across all dimensions. 
For sparse regression (with both continuous and discrete variables),
we measure the posterior approximation quality and sampling efficiency for the discrete
and continuous posterior marginals separately. For the continuous variables, we use the same metrics
as for the other experiments. For the discrete variables, we measure posterior approximation quality via the 
Jensen-Shannon divergence---which accounts for the possiblity of differing support in empirical distribution
approximations---and efficiency via the ESS computed for the fraction of correct feature inclusion indicators.
For both ESS computations, we use the bulk-ESS formula in \cite{vehtari2021rank}.

For all coreset methods that use an MCMC kernel, we use the hit-and-run slice sampler with doubling 
\citep{belisle1993hit,neal2003slice} for linear and logistic regressions, the univariate slice sampler with doubling 
(\citealp[Fig.~4]{neal2003slice}) applied to each dimension for the more challenging Poisson regression, 
and the Gibbs sampler developed by \cite{george1993variable} for sparse regression. 
For \texttt{SGLD-CV} and \texttt{SGHMC-CV}, we set the subsampling size to $500$. For \texttt{Austerity} and 
\texttt{Confidence}, we set the accept-reject decision threshold to $0.05$.

All experiments were performed on 
the UBC ARC Sockeye cluster. 
Each algorithm was run on 8 single-threaded cores of a 2.1GHz Intel Xeon Gold 6130 processor with 32GB memory. 
Code for these experiments is available at
\url{https://github.com/NaitongChen/coreset-mcmc-experiments}.
More experimental details 
and additional plots are in \cref{sec:expt,sec:additional_results}.

\captionsetup[subfigure]{labelformat=empty}
\begin{figure*}[h!]
    \centering 
\begin{subfigure}[b]{.3\textwidth} 
    \scalebox{1}{\includegraphics[width=\textwidth]{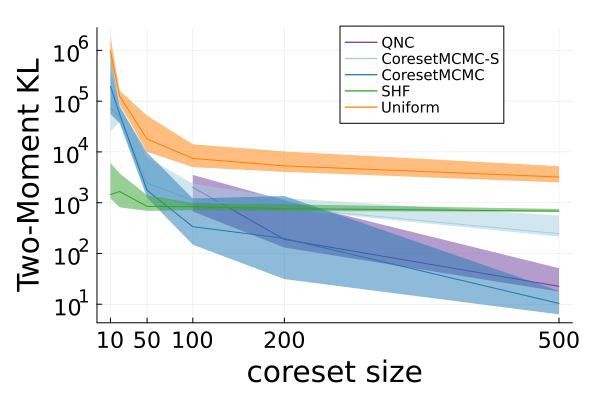}}
    \caption{(a) linear regression\label{fig:kl_coreset_lin}}
\end{subfigure}
\hfill
\centering
\begin{subfigure}[b]{0.3\textwidth}
    \scalebox{1}{\includegraphics[width=\textwidth]{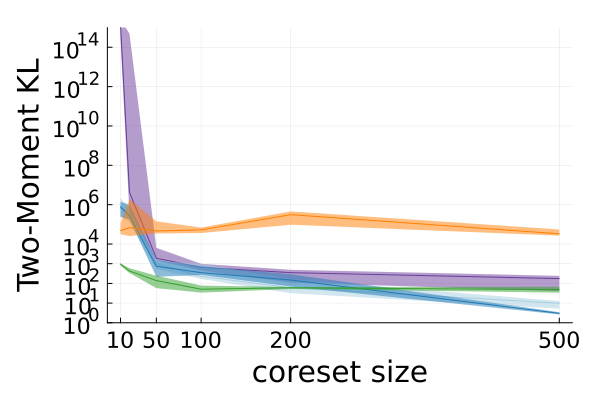}}
    \caption{(b) logistic regression\label{fig:kl_coreset_log}}
\end{subfigure}
\hfill
\centering 
\begin{subfigure}[b]{.3\textwidth} 
    \scalebox{1}{\includegraphics[width=\textwidth]{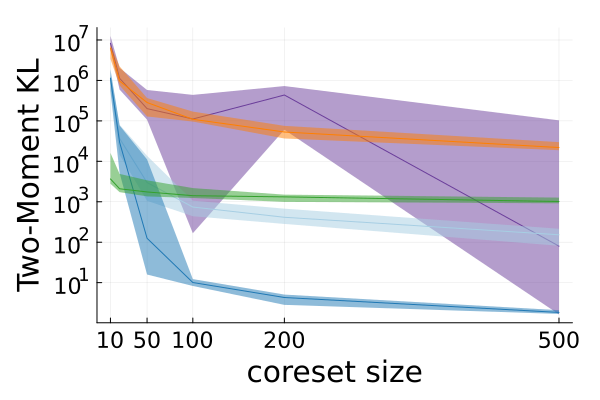}}
    \caption{(c) Poisson regression\label{fig:kl_coreset_poi}}
\end{subfigure}
\centering
\begin{subfigure}[b]{0.3\textwidth}
    \scalebox{1}{\includegraphics[width=\textwidth]{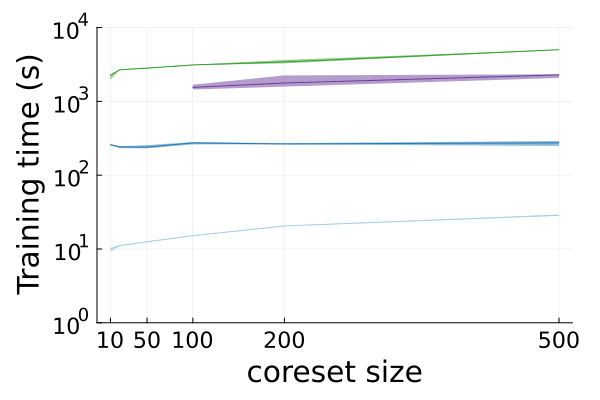}}
    \caption{(d) linear regression\label{fig:lin_train_coreset}}
\end{subfigure}
\hfill
\centering
\begin{subfigure}[b]{0.3\textwidth}
    \scalebox{1}{\includegraphics[width=\textwidth]{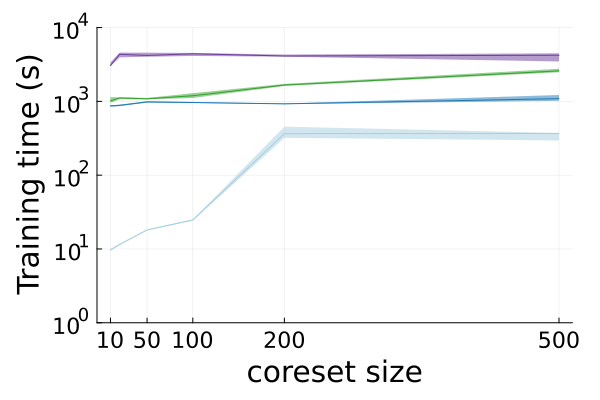}}
    \caption{(e) logistic regression\label{fig:log_train_coreset}}
\end{subfigure}
\hfill
\centering
\begin{subfigure}[b]{0.3\textwidth}
    \scalebox{1}{\includegraphics[width=\textwidth]{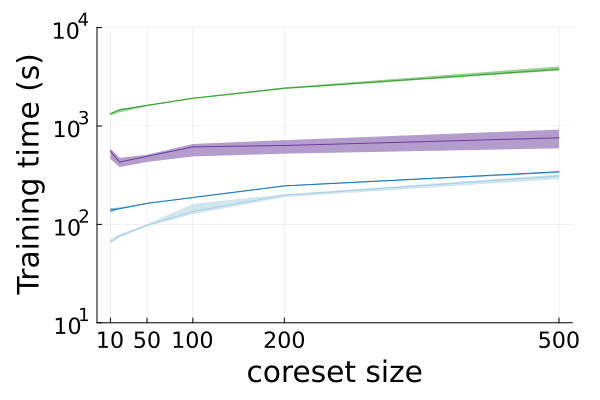}}
    \caption{(f) Poisson regression\label{fig:poi_train_coreset}}
\end{subfigure}
\hfill
\centering
\begin{subfigure}[b]{0.3\textwidth}
    \scalebox{1}{\includegraphics[width=\textwidth]{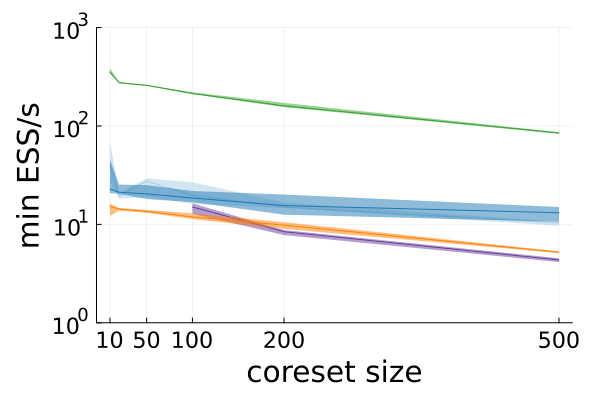}}
    \caption{(g) linear regression\label{fig:lin_ess_coreset}}
\end{subfigure}
\hfill
\centering
\begin{subfigure}[b]{0.3\textwidth}
    \scalebox{1}{\includegraphics[width=\textwidth]{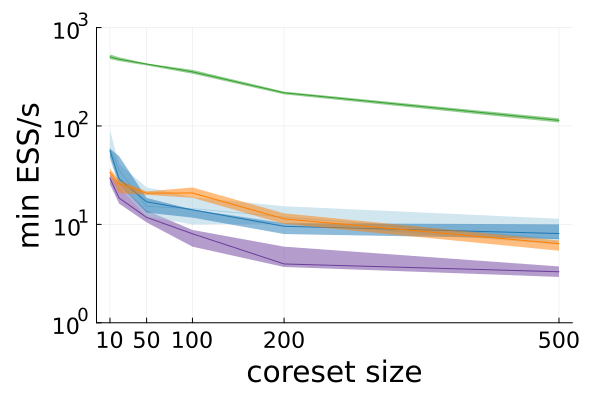}}
    \caption{(h) logistic regression\label{fig:log_ess_coreset}}
\end{subfigure}
\hfill
\centering
\begin{subfigure}[b]{0.3\textwidth}
    \scalebox{1}{\includegraphics[width=\textwidth]{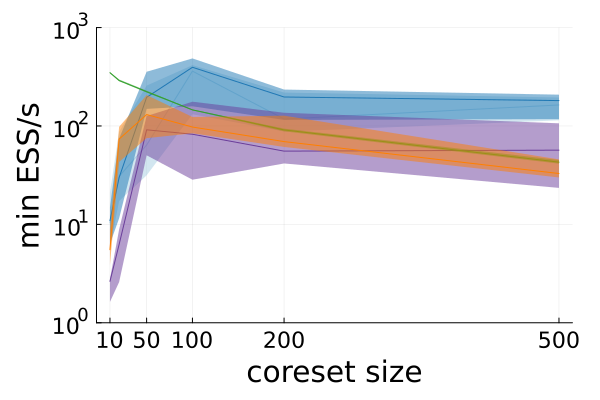}}
    \caption{(i) Poisson regression\label{fig:poi_ess_coreset}}
\end{subfigure}
\hfill
\caption{Comparison of coreset methods on real data examples. \cref{fig:kl_coreset_lin,fig:kl_coreset_log,fig:kl_coreset_poi} show
posterior approximation quality via the two-moment KL, \cref{fig:lin_train_coreset,fig:log_train_coreset,fig:poi_train_coreset} show
training time, and \cref{fig:lin_ess_coreset,fig:log_ess_coreset,fig:poi_ess_coreset} show
 sampling efficiency via the min. ESS per second. 
The lines indicate the median, and error regions indicate 25$^\text{th}$ to 75$^\text{th}$ percentile from 10 runs.}
\label{fig:coreset}
\end{figure*}

\subsection{Comparison of Coreset Methods}

\cref{fig:coreset} displays the comparison of Coreset MCMC with past coreset construction methods on the three real data experiments
over various coreset sizes $M$.
As shown in \cref{fig:kl_coreset_lin,fig:kl_coreset_log,fig:kl_coreset_poi},
the \texttt{Unif} baseline, although not requiring any training and yielding a competitive sampling efficiency,
generally provides poor quality coresets with an order of magnitude
higher two-moment KL nearly uniformly across all coreset sizes than all other methods. 
\texttt{QNC} generates competitive posterior approximations for both linear and
logistic regression, but a significantly worse approximation in the more challenging Poisson regression problem.
Its approximations are also less reliable in general; the instability is particularly evident in
linear regression when the coreset size is small---where \texttt{QNC} produces
\texttt{NaN} values even after multiple attempts at tuning---and in the poisson regression example.  
\texttt{SHF} generally provides high quality approximations for small coreset sizes, 
but they do not improve with increasing coreset size. We conjecture that this is due
to the quasi-refreshment steps limiting the expressiveness of the variational
family used in \texttt{SHF}.

\cref{fig:lin_train_coreset,fig:log_train_coreset,fig:poi_train_coreset} show that 
both \texttt{QNC} and \texttt{SHF} take between 2 to 10 times longer to train than both variants of Coreset MCMC. 
\texttt{QNC} takes longer than \texttt{CoresetMCMC} to train because it runs a full MCMC procedure at 
each optimization iteration, and computes a quasi-Newton update that has time complexity $\mathcal{O}(M^3)$;
compare to \texttt{CoresetMCMC}, which has updates that require only $\mathcal{O}(M)$ time.
\texttt{SHF} takes longer than \texttt{CoresetMCMC} to train because it requires taking the gradient over the 
entire flow to obtain updates on the coreset weights, refreshment parameters, and step sizes. 
Once trained, \cref{fig:lin_ess_coreset,fig:log_ess_coreset,fig:poi_ess_coreset} show that all coreset methods provide
a similar minimum ESS per second except for \texttt{SHF}, which is a variational methods that provides \iid draws. 

\captionsetup[subfigure]{labelformat=empty}
\begin{figure*}[h!]
    \centering
\begin{subfigure}[b]{0.19\textwidth}
    \scalebox{1}{\includegraphics[width=\textwidth]{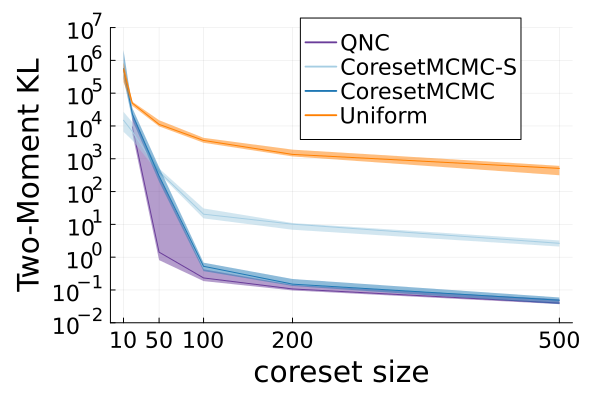}}
    \caption{(a) quality (cont.) \label{fig:kl_coreset_sp}}
\end{subfigure}
\hfill
\centering
\begin{subfigure}[b]{0.19\textwidth}
    \scalebox{1}{\includegraphics[width=\textwidth]{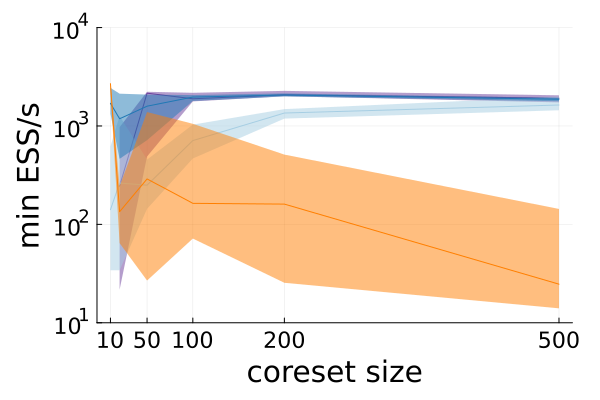}}
    \caption{(b) efficiency (cont.) \label{fig:sp_ess_coreset}}
\end{subfigure}
\hfill
\centering
\begin{subfigure}[b]{0.19\textwidth}
    \scalebox{1}{\includegraphics[width=\textwidth]{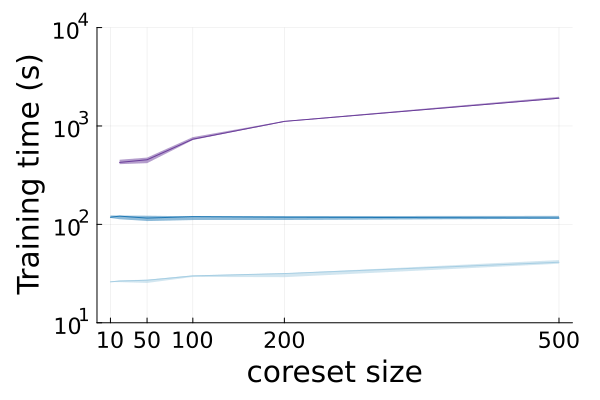}}
    \caption{(c) training time \label{fig:sp_train_coreset}}
\end{subfigure}
\hfill
\centering
\begin{subfigure}[b]{0.19\textwidth}
    \scalebox{1}{\includegraphics[width=\textwidth]{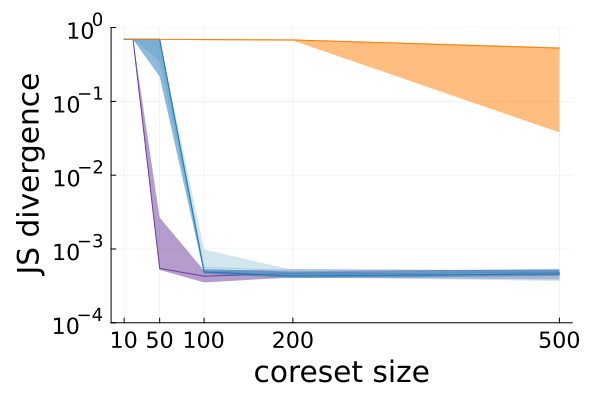}}
    \caption{(d) quality (disc.) \label{fig:kld_coreset_sp}}
\end{subfigure}
\hfill
\centering
\begin{subfigure}[b]{0.19\textwidth}
    \scalebox{1}{\includegraphics[width=\textwidth]{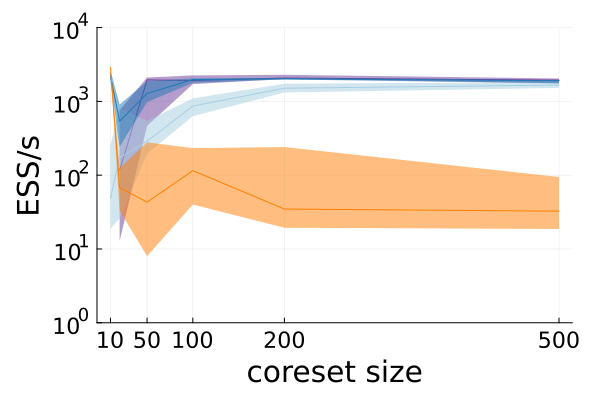}}
    \caption{(e) efficiency (disc.) \label{fig:sp_essd_coreset}}
\end{subfigure}
\hfill
\caption{Comparison of coreset methods on sparse regression. 
\cref{fig:kl_coreset_sp,fig:sp_ess_coreset} show posterior 
approximation quality and sampling efficiency across the continuous components via the two-moment KL and the 
min. ESS per second.
\cref{fig:kld_coreset_sp,fig:sp_essd_coreset} show those across the discrete components via the Jensen-Shannon divergence and ESS per second. 
\cref{fig:sp_train_coreset} shows training time in seconds.
The lines indicate the median, and error regions indicate 25$^\text{th}$ to 75$^\text{th}$ 
percentile from 10 runs.}
\label{fig:sp_coreset}
\end{figure*}

\cref{fig:sp_coreset} shows the same comparison of coreset methods on sparse regression. 
Note that \texttt{SHF} is not applicable here due to the inclusion of discrete variables, and hence 
is excluded from this experiment.
We see that the trends in these plots resemble those from the three real data experiments: \texttt{Unif} 
produces worse quality coresets than \texttt{CoresetMCMC} and \texttt{QNC}, and \texttt{QNC} requires 
much longer training times than \texttt{CoresetMCMC}.

Comparing between the two variants of Coreset MCMC methods, it is clear
that by including the full dataset when estimating the gradient, we obtain
generally higher quality posterior approximations in terms of the two-moment KL.
In general, we recommend setting $S$ in \cref{eq:gradest} such that 
gradient updates require similar time as generating the next MCMC state,
to balance time spent on simulating the Markov chains and estimating KL gradients.

\captionsetup[subfigure]{labelformat=empty}
\begin{figure*}[t!]
\centering
\begin{subfigure}[b]{0.3\textwidth}
    \scalebox{1}{\includegraphics[width=\textwidth]{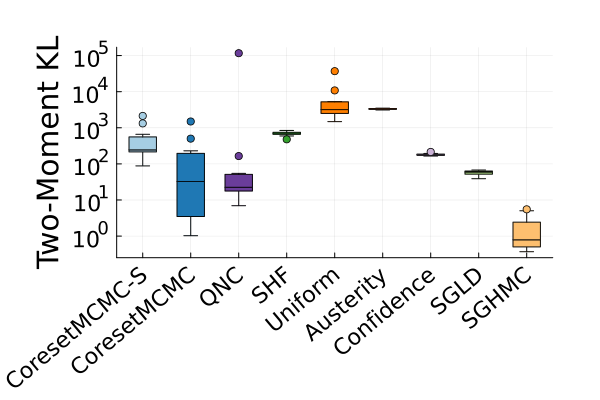}}
    \caption{(a) linear regression \label{fig:kl_all_lin}}
\end{subfigure}
\hfill
\centering
\begin{subfigure}[b]{0.3\textwidth}
    \scalebox{1}{\includegraphics[width=\textwidth]{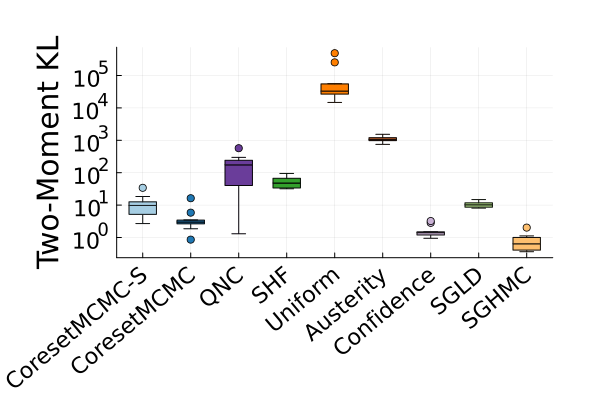}}
    \caption{(b) logistic regression \label{fig:kl_all_log}}
\end{subfigure}
\hfill
\centering
\begin{subfigure}[b]{0.3\textwidth}
    \scalebox{1}{\includegraphics[width=\textwidth]{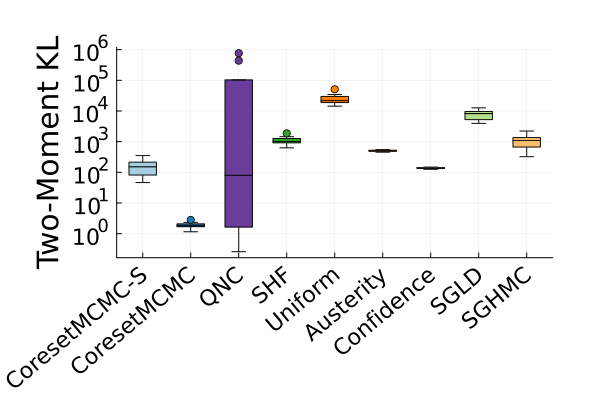}}
    \caption{(c) Poisson regression \label{fig:kl_all_poi}}
\end{subfigure}
\hfill
\centering
\begin{subfigure}[b]{0.3\textwidth}
    \scalebox{1}{\includegraphics[width=\textwidth]{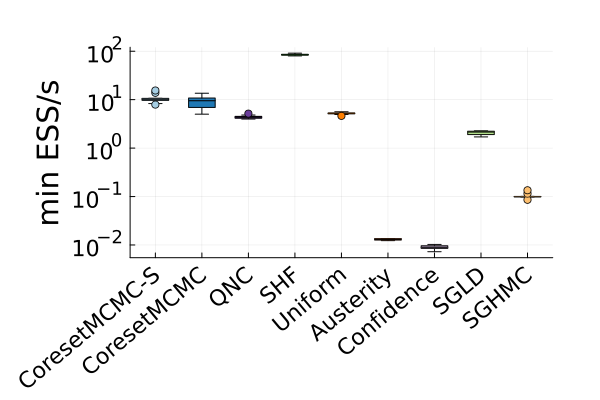}}
    \caption{(d) linear regression \label{fig:lin_ess_all}}
\end{subfigure}
\hfill
\centering
\begin{subfigure}[b]{0.3\textwidth}
    \scalebox{1}{\includegraphics[width=\textwidth]{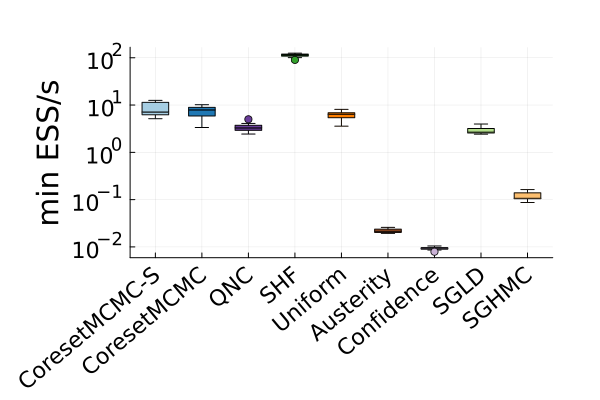}}
    \caption{(e) logistic regression \label{fig:log_ess_all}}
\end{subfigure}
\hfill
\centering
\begin{subfigure}[b]{0.3\textwidth}
    \scalebox{1}{\includegraphics[width=\textwidth]{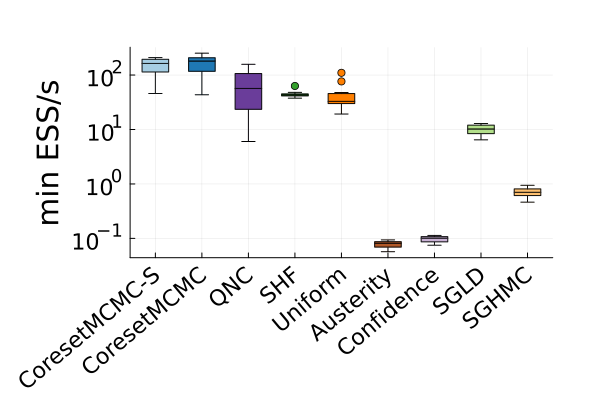}}
    \caption{(f) Poisson regression \label{fig:poi_ess_all}}
\end{subfigure}
\hfill

\caption{Comparison of coreset and subsampling MCMC methods on real data examples.
\cref{fig:kl_all_lin,fig:kl_all_log,fig:kl_all_poi} show posterior approximation
quality via the two-moment KL, and \cref{fig:lin_ess_all,fig:log_ess_all,fig:poi_ess_all}
show sampling efficiency via the min. ESS per second.
The boxplots indicate the median, 25$^\text{th}$, and 75$^\text{th}$ percentiles from 10 runs.}
\label{fig:all}
\end{figure*}

\captionsetup[subfigure]{labelformat=empty}
\begin{figure*}[t!]
    \centering
    \begin{subfigure}[b]{0.24\textwidth}
        \scalebox{1}{\includegraphics[width=\textwidth]{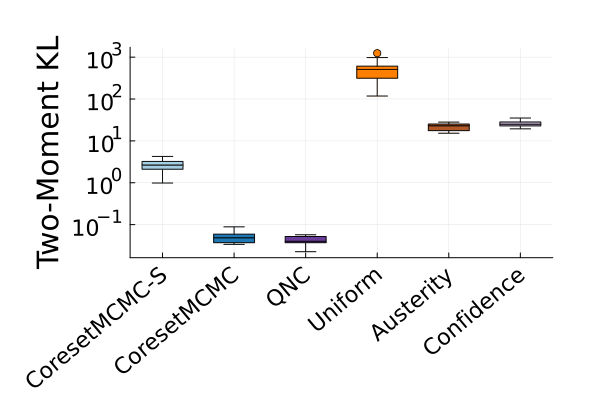}}
        \caption{(a) quality (cont.) \label{fig:kl_all_sp}}
    \end{subfigure}
    \hfill
    \centering
    \begin{subfigure}[b]{0.24\textwidth}
        \scalebox{1}{\includegraphics[width=\textwidth]{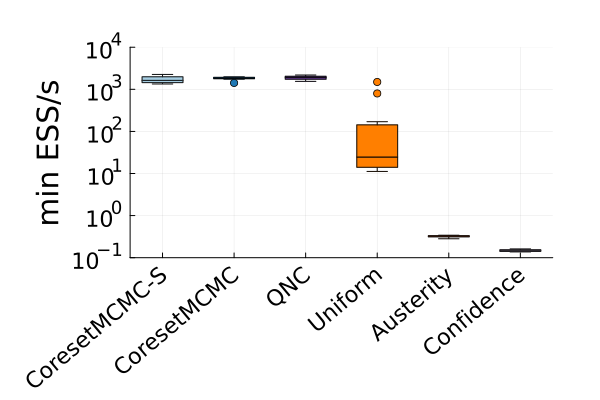}}
        \caption{(b) efficiency (cont.) \label{fig:sp_ess_all}}
    \end{subfigure}
    \hfill
    \centering
    \begin{subfigure}[b]{0.24\textwidth}
        \scalebox{1}{\includegraphics[width=\textwidth]{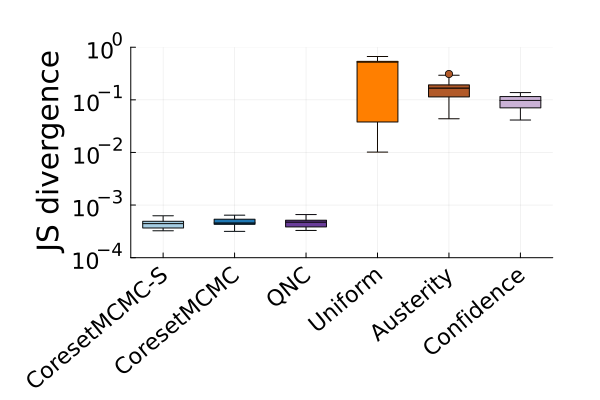}}
        \caption{(c) quality (disc.) \label{fig:kl_disc_sp}}
    \end{subfigure}
    \hfill
    \centering
    \begin{subfigure}[b]{0.24\textwidth}
        \scalebox{1}{\includegraphics[width=\textwidth]{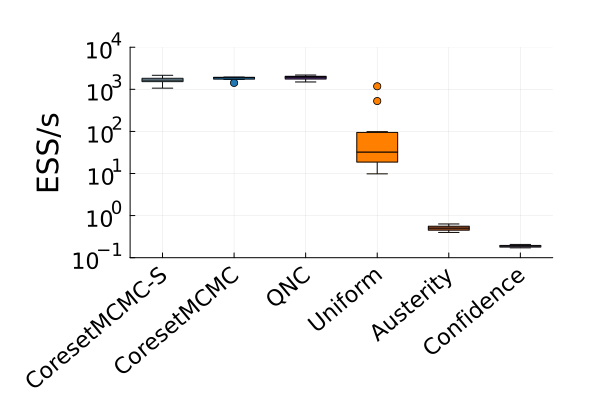}}
        \caption{(d) efficiency (disc.) \label{fig:sp_ess_disc}}
    \end{subfigure}

\caption{Comparison of coreset and subsampling MCMC methods on sparse regression. 
\cref{fig:kl_all_sp,fig:sp_ess_all} show posterior approximation quality and sampling efficiency 
across the continuous components via the two-moment KL and the min. ESS per second. 
\cref{fig:kl_disc_sp,fig:sp_ess_disc} show those across the discrete components via the Jensen-Shannon divergence and ESS per second.
The boxplots indicate the median, 25$^\text{th}$, and 75$^\text{th}$ percentiles from 10 runs.}
\label{fig:sp_all}
\end{figure*}

\subsection{Comparison with Subsampling MCMC}
\cref{fig:all,fig:sp_all} display the comparison between all coreset methods and all subsampling MCMC methods, 
on the three real data experiments and the synthetic sparse regression experiment. Here we fix the coreset size to $M=500$.
On the easier linear and logistic regression problems (i.e., with more Gaussian-like posteriors), we see that \texttt{CoresetMCMC} provides
approximations competitive, and sometimes better, than the other methods 
except for \texttt{SGHMC-CV}. However, we note that stochastic gradient MCMC methods depends heavily on the quality of the control variate. This 
is illustrated in the more challenging Poisson regression example; in this example, the control variate construction 
struggled to identify the mode of the posterior, causing resulting approximation quality to suffer. 
Furthermore, stochastic gradient MCMC methods are limited to models with only continuous variables, whereas 
\texttt{CoresetMCMC}, as illustrated in \cref{fig:sp_all}, applies to and performs well on models with both 
continuous and discrete variables.

In terms of sampling efficiency,
\texttt{CoresetMCMC} and \texttt{CoresetMCMC-S} are uniformly 
better than other subsampling MCMC methods. Specifically, we get roughly $2$ orders of magnitude higher ESS per second 
in \texttt{CoresetMCMC} than \texttt{SGHMC-CV}. This means that \texttt{CoresetMCMC} can eventually catch up despite 
the initial training time. As an example, in the linear 
regression case, although it takes \texttt{CoresetMCMC} roughly $500$ seconds to train, it is only enough time for 
\texttt{SGHMC-CV} to obtain roughly $50$ effective samples. This deficit can be recovered by \texttt{CoresetMCMC} after 
$5$ seconds once it stops adapting the coreset weights.

\subsection{Tuning Difficulty}

Although not reflected in the results in \cref{fig:coreset,fig:sp_coreset,fig:all,fig:sp_all}, tuning difficulty
was a key differentiator in usability among the methods tested. Compared with other coreset methods,
we found \texttt{CoresetMCMC} very straightforward to tune, as it just requires setting a learning rate.
For \texttt{QNC}, in contrast, one needs to pick the number of MCMC samples used to estimate the gradient, 
the optimization step size, the number of iterations in which to perform a line search on the step size, and the number of 
weight optimization iterations. We note that the recommendation of setting the optimization step size to $1$ from \cite{naik2022fast} may cause 
the coreset weights to become unstable; we instead set this value to the step size at the last iteration where 
a line search was performed. Despite this modification, we can see from our results that \texttt{QNC} is still less 
stable than other coreset methods, and as such required significantly more manual tuning effort and time than \texttt{CoresetMCMC}.
For \texttt{SHF}, much of the tuning effort goes into specifying the flow architecture: one must select 
the number of quasi-refreshment steps and the number of leapfrog steps in between quasi-refreshments. As discussed earlier, 
these tuning parameters determine the  expressiveness of the variational family, and so one needs to balance having a highly expressive variational family 
without it being too complex to train within reasonable time.

For the stochastic gradient MCMC methods with control variates, the key difficulty (beyond tuning the subsample size) was to
balance the effort spent constructing a good control variate (i.e., finding a reasonable estimate of the posterior mode) versus
the effort spent using the control variate in sampling. For \texttt{SGHMC-CV}, one also needs to specify the number of 
leapfrog steps and the leapfrog step size. These two parameters are as complicated to tune as in
Hamiltonian Monte Carlo.

Finally, both \texttt{Austerity} and \texttt{Confidence} require the user to input a threshold that determines the quality of 
the accept-reject decision approximation. This quantity does not directly influence the sampling efficiency as 
subsample size would. As a result, if one has a clear computation time budget, it may be more intricate to tune these 
methods well.

%% file: conclusion.tex
\section{\uppercase{Conclusion}}
\label{sec:conclusion}
This paper introduced Coreset MCMC, a novel coreset construction method that is simple to implement and tune compared 
with previous coreset methods. Coreset MCMC involves interleaving simulating states from several independent Markov chains 
targeting the coreset posterior with updating the coreset weights using these same draws. Theoretical results demonstrated that our method is 
asymptotically exact, assuming there exists a coreset whose corresponding posterior approximation has KL 0 to the 
true posterior, as well as other technical conditions. Finally, empirical results demonstrated that our method provides higher quality posterior
approximations compared with other coreset methods, and improved sampling efficiency compared with subsampling MCMC methods.

%% file: proofs.tex
\section{\uppercase{Proofs}}\label{sec:proofs}
\bprfof{\cref{thm:converge_subsample}}
Define
\[
H_t \defas \frac{1}{K-1}\sum_{k=1}^K
\bbmat \bar\ell_1(\theta_{tk})\\ \vdots \\ \bar\ell_M(\theta_{tk})\ebmat
\bbmat \bar\ell_1(\theta_{tk})\\ \vdots \\ \bar\ell_N(\theta_{tk})\ebmat^\top
\in \reals^{M\times N}.
\]
Note that $H_t$ is similar to $G_t$ except that there are $N$ columns.
Define $s_t \in \reals^N$ such that $\forall j \in \fcS_t, s_{tj} = \frac{N}{S}$ and $0$ otherwise.
The subsampled gradient estimate can then be written in the form
\[
g(w_t, \theta_t, \fcS_t) &= G_t(w_t - w^\star) + H_t(1 - s_t).
\]
 We apply the projected gradient update to get
\[
\|w_{t+1} - w^\star\|^2
&=\|\proj_{\fcW}\lt(w_t - \gamma_t G_t(w_t - w^\star) - \gamma_t H_t(1 - s_t)\rt) - w^\star\|^2\\
&=\|\proj_{\fcW}\lt(w_t - \gamma_t G_t(w_t - w^\star) - \gamma_t H_t(1 - s_t)\rt) - \proj_{\fcW}w^\star\|^2\\
&\leq\|w_t - \gamma_t G_t(w_t - w^\star) - \gamma_t H_t(1 - s_t) - w^\star\|^2\\
&=\|(I- \gamma_t G_t)(w_t - w^\star) - \gamma_t H_t(1 - s_t)\|^2.
\]
The second equality follows because $w^\star \in \fcW$ by assumption. 
The inequality follows because $\fcW$ is convex and closed
by assumption,
and hence $\proj_{\fcW}$ is a contraction.
Therefore,
\[
\|w_{t+1} - w^\star\|^2 &\leq \|(I-\gamma_t G_t)(w_t-w^\star) - \gamma_t H_t(1-s_t)\|^2.
\]
Expand the right hand side above and take the expectation, we get
\[
&\ex \|w_{t+1} - w^\star\|^2 \\
&\leq \ex \lt [ (w_t-w^\star)^\top (I-\gamma_t G_t)^\top(I-\gamma_t G_t)(w_t-w^\star) \rt ] + \gamma^2_t \ex \lt[(1-s_t)^\top H_t^\top H_t(1-s_t)\rt] \\
&\quad - 2\gamma_t \ex \lt[ (w_t - w^\star)^\top (I - \gamma_t G_t)^\top H_t(1-s_t)  \rt]. \label{eq:ex_update}
\]
Now use the tower property to show that the last term above is $0$:
\[
\ex \lt[ (w_t - w^\star)^\top (I - \gamma_t G_t)^\top H_t(1-s_t) \rt] 
&= \ex \lt[ \ex \lt[ (w_t - w^\star)^\top (I - \gamma_t G_t)^\top H_t(1-s_t) \given w_t, \theta_t \rt] \rt]\\
&= \ex \lt[ (w_t - w^\star)^\top (I - \gamma_t G_t)^\top H_t \ex \lt [(1-s_t) \given w_t, \theta_t \rt] \rt]\\
&= 0,
\]
where the second line follows by noting $\ex\lt[1-s_t\given w_t,\theta_t\rt] = \ex \lt[ 1 - s_t \rt] = 0$ due to the unbiased subsampling.
We now focus on the second term on the right-hand side of \cref{eq:ex_update}.
Again note that $\ex\lt[s_t \given \theta_t\rt] = \ex\lt[s_t\rt] = 1$, so
\[
&\ex \lt[(1-s_t)^\top H_t^\top H_t(1-s_t) \given \theta_t\rt]\\
&= - 1^\top H_t^\top H_t1 + \ex \lt[s_t^\top H_t^\top H_ts_t\given \theta_t\rt]\\
&= \ex \lt[\sum_{n,n'} s_{tn}s_{tn'}(H_t^\top H_t)_{nn'}\given \theta_t\rt] - \sum_{n,n'}(H_t^\top H_t)_{nn'}\\
&= \frac{N}{S}\lt(\sum_n\lt(1-\frac{S}{N}\rt) (H_t^\top H_t)_{nn} + \sum_{n\neq n'}\lt(\frac{S-1}{N-1}-\frac{S}{N}\rt) (H_t^\top H_t)_{nn'}\rt)\\
&= \frac{N}{S}\lt(\sum_n\lt(\frac{N-S}{N}\rt) (H_t^\top H_t)_{nn} - \sum_{n\neq n'}\frac{N-S}{N(N-1)} (H_t^\top H_t)_{nn'}\rt)\\
&= \frac{N(N-S)}{S(N-1)}\lt(\sum_n(H_t^\top H_t)_{nn} - \frac{1}{N}\sum_{n,n'}(H_t^\top H_t)_{nn'}\rt)\\
&= \frac{N^2(N-S)}{S(N-1)(K-1)^2}\sum_{k,k'}\lt(\sum_m \bar\ell_m(\theta_{tk})\bar\ell_m(\theta_{tk'}) \rt)\lt(
\frac{1}{N}\sum_n \bar\ell_n(\theta_{tk})\bar\ell_n(\theta_{tk'})
- \frac{1}{N^2}\sum_{n,n'}\bar\ell_n(\theta_{tk})\bar\ell_{n'}(\theta_{tk'})
\rt)\\
&= \! \frac{N^2(N\!-\!S)}{S(N\!-\!1)(K\!-\!1)^2}\!\sum_{k,k'}\!\lt(\sum_m \bar\ell_m(\theta_{tk})\bar\ell_m(\theta_{tk'}) \rt)\!\lt(
	\!\frac{1}{N}\!\sum_n\!\lt(\!\bar\ell_n(\theta_{tk}) - \frac{1}{N}\sum_{n'}\bar\ell_{n'}(\theta_{tk})\!\rt)\!
\lt(\!\bar\ell_n(\theta_{tk'}) - \frac{1}{N}\sum_{n'}\bar\ell_{n'}(\theta_{tk'})\!\rt)
\rt) \\
&= \frac{N^2M(N-S)}{S(N-1)} V_t,
\]
where $V_t$ is as defined in \cref{assump:noisebounded}.
Therefore,
\[
\ex \lt[(1-s_t)^\top H_t^\top H_t(1-s_t)\rt] &= \ex\lt[\ex \lt[(1-s_t)^\top H_t^\top H_t(1-s_t) \given \theta_t\rt] \rt]\\
&= \frac{N^2M(N-S)}{S(N-1)} \ex\lt[ V_t \rt].
\]
Given the previous two steps, \cref{eq:ex_update} becomes
\[
\ex \|w_{t+1} - w^\star\|^2 &\leq 
\ex \lt[ (w_t-w^\star)^\top (I-\gamma_t G_t)^\top (I-\gamma_t G_t)(w_t-w^\star) \rt] + \gamma^2_t \frac{N^2M(N-S)}{S(N-1)} \ex\lt[V_t\rt].
\]
By \cref{assump:mixing,assump:bounded},  we can rewrite the first term above as 
\[
&\ex \lt[ (w_t-w^\star)^\top (I-\gamma_t G_t)^\top (I-\gamma_t G_t)(w_t-w^\star) \rt]\\
&= \ex \lt[ \ex \lt[ (w_t-w^\star)^\top (I-\gamma_t G_t)^\top (I-\gamma_t G_t)(w_t-w^\star) \given w_t, \theta_{t-1} \rt] \rt]\\
&= \ex \lt[ (w_t-w^\star)^\top \lt(I-2\gamma_t \ex\lt[G_t\given w_t, \theta_{t-1}\rt] + \gamma_t^2 \ex\lt[G_t^\top G_t\given w_t, \theta_{t-1}\rt]\rt)(w_t-w^\star) \rt]\\
&\leq \ex\lt[ (w_t-w^\star)^\top \lt(I-2\gamma_t \lambda I + \gamma_t^2 \bar\lambda I\rt)(w_t-w^\star)\rt].
\]
By \cref{assump:noisebounded}, we have
\[
\ex\lt[V_t\rt] = \ex\lt[ \ex\lt[ V_t \given w_t, \theta_{t-1} \rt] \rt] \leq V.
\]
Therefore

\[
\ex \|w_{t+1} - w^\star\|^2
&\leq\ex\lt[ (w_t-w^\star_t)^\top \lt(I-2\gamma_t \lambda I + \gamma_t^2 \bar\lambda I\rt)(w_t-w^\star_t)\rt] + \gamma^2_t \frac{N^2M(N-S)}{S(N-1)}V\\
&= \ex \lt [ \lt(1 - 2\gamma_t \lambda  + \gamma_t^2 \bar\lambda \rt) \| w_t-w^\star\|^2\rt] + \gamma^2_t \frac{N^2M(N-S)}{S(N-1)}V\\
&= \lt(1 - 2\gamma_t \lambda  + \gamma_t^2 \bar\lambda \rt) \ex \| w_t-w^\star\|^2 + \gamma^2_t \frac{N^2M(N-S)}{S(N-1)}V.
\]
Now let $\gamma = \lambda /\overline{\lambda} > 0$, and suppose $\sup_t \gamma_t \leq \gamma$. Then
\[
\ex \|w_{t+1} - w^\star\|^2 
&\leq \lt(1 - \gamma_t \lambda  \rt) \ex \| w_t-w^\star\|^2 + \gamma^2_t \frac{N^2M(N-S)}{S(N-1)}V\\
&\leq e^{-\gamma_t \lambda } \ex \| w_t-w^\star\|^2 + \gamma^2_t \frac{N^2M(N-S)}{S(N-1)}V.
\]
Then expand the recursion to obtain the stated bound:
\[
\ex \|w_{t} - w^\star\|^2 
&\leq e^{-\lambda \sum_{\tau=0}^{t-1}\gamma_\tau}\|w_0-w^\star\|^2 + \frac{N^2M(N-S) V}{S(N-1)}\sum_{\tau=0}^{t-1}\gamma_\tau^2 e^{-\lambda \sum_{u=\tau+1}^{t-1}\gamma_u}.\label{eq:simple}
\]
Convergence of the first term to $0$ as $t\to\infty$ occurs as long as $\sum_{t=0}^\infty \gamma_t = \infty$.
It remains to show that the second sum term converges to $0$ as $t\to\infty$ when additionally $\gamma_t \to 0$.
Note that since $\sum_{t=0}^\infty \gamma_t = \infty$, 
\[
\gamma_0^2 e^{-\lambda \sum_{u=1}^{t-1}\gamma_u} \to 0, \qquad t\to\infty,
\]
and so the $0$-index term in the sum converges to 0. Therefore we can focus on the sum from indices 1 to $t-1$.
Note that since $\gamma_t$ is a monotone decreasing sequence, it can be expanded to a monotone decreasing continuous 
real function through linear interpolation, so
\[
\sum_{u=\tau+1}^{t-1}\gamma_u \geq \int_{\tau+1}^{t} \gamma_u \dee u,
\]
and thus we have
\[
\sum_{\tau=1}^{t-1}\gamma_\tau^2 e^{-\lambda \sum_{u=\tau+1}^{t}\gamma_u}
\leq 
\sum_{\tau=1}^{t-1}\gamma_\tau^2 e^{-\lambda \int_{\tau+1}^{t} \gamma_u \dee u}.
\]
Next we split the sum into two sums separated by some arbitrary index $1 < b < t$, 
\[
\sum_{\tau=1}^{t-1}\gamma_\tau^2 e^{-\lambda \int_{\tau+1}^{t} \gamma_u \dee u}
=
\sum_{\tau=1}^{b-1}\gamma_\tau^2 e^{-\lambda \int_{\tau+1}^{t} \gamma_u \dee u}
+
\sum_{\tau=b}^{t-1}\gamma_\tau^2 e^{-\lambda \int_{\tau+1}^{t} \gamma_u \dee u}. \label{eq:halfbound}
\]

We now bound each of the above two terms. The second term in \cref{eq:halfbound} above can be bounded by a sum with a monotone decreasing summand, which can then be bounded with an integral:
\[
\sum_{\tau=b}^{t-1}\gamma_\tau^2 e^{-\lambda \int_{\tau+1}^{t} \gamma_u \dee u}
&\leq \sum_{\tau=b}^{t-1}\gamma_\tau^2 \leq \int_b^t \gamma^2_{\tau-1}\dee \tau.
\]
The first sum term in \cref{eq:halfbound} has a summand that is a product of a monotone decreasing and a monotone increasing sequence. 
We can split this sum further by isolating the summand at $\tau = 0$:
\[
\sum_{\tau=1}^{b-1}\gamma_\tau^2 e^{-\lambda \int_{\tau+1}^{t} \gamma_u \dee u}
&=  \sum_{\tau=1}^{b-1}\gamma_\tau^2 e^{-\lambda \int_{\tau+1}^{t} \gamma_u \dee u}\\
&\leq  
\int_1^{b}\gamma_{\tau-1}^2 e^{-\lambda \int_{\tau+1}^{t} \gamma_u \dee u}\dee \tau\\
&= 
e^{-\lambda \int_0^t \gamma_u\dee u} \int_1^{b}\gamma_{\tau-1}^2 e^{\lambda \int_{0}^{\tau+1} \gamma_u \dee u}\dee \tau\\
&= 
e^{-\lambda \int_0^t \gamma_u\dee u} \int_1^{b}\gamma_{\tau-1}^2 e^{\lambda \int_{0}^{\tau-1} \gamma_u \dee u}e^{\lambda \int_{\tau-1}^{\tau+1} \gamma_u \dee u}\dee \tau\\
&\leq
e^{-\lambda \int_0^t \gamma_u\dee u} \int_1^{b}\gamma_{\tau-1}^2 e^{\lambda \int_{0}^{\tau-1} \gamma_u \dee u}e^{\lambda \int_{0}^{2} \gamma_u \dee u}\dee \tau\\
&\leq e^{-\lambda \int_0^t \gamma_u\dee u + \lambda 2\gamma} \int_1^{b}\gamma_{\tau-1}^2 e^{\lambda \int_{0}^{\tau-1} \gamma_u \dee u}\dee \tau\\
&\leq \gamma e^{-\lambda \int_0^t \gamma_u\dee u + \lambda 2\gamma} \int_1^{b}\gamma_{\tau-1} e^{\lambda \int_{0}^{\tau-1} \gamma_u \dee u}\dee \tau,
\]
where all inequalities follow by noting that $\gamma_{t}$ is monotone decreasing.
Now apply the change of variables $x = \int_{0}^{\tau-1} \gamma_u \dee u$ with
$\dee x = \gamma_{\tau-1} \dee \tau$ to get
\[
\int_1^{b}\gamma_{\tau-1} e^{\lambda \int_{0}^{\tau-1} \gamma_u \dee u}\dee \tau
= \int_0^{\int_0^{b-1}\gamma_u \dee u} e^{\lambda x}\dee x
= \frac{e^{\lambda \int_0^{b-1}\gamma_u \dee u} - 1}{\lambda}
\leq \lambda^{-1}e^{\lambda \int_0^{b}\gamma_u \dee u}.
\]
Therefore the first term in \cref{eq:halfbound} becomes
\[
\sum_{\tau=1}^{b-1}\gamma_\tau^2 e^{-\lambda \int_{\tau+1}^{t} \gamma_u \dee u}
&\leq  \gamma \lambda^{-1} e^{-\lambda \int_0^t \gamma_u\dee u + \lambda 2\gamma + \lambda \int_0^{b}\gamma_u \dee u}
=  \gamma \lambda^{-1} e^{2\lambda \gamma -\lambda \int_b^t \gamma_u\dee u }.
\]
Therefore we have that 
\[
\sum_{\tau=1}^{t-1}\gamma_\tau^2 e^{-\lambda \int_{\tau+1}^{t} \gamma_u \dee u} 
\leq  \int_b^t \gamma^2_{\tau-1}\dee \tau + \gamma \lambda^{-1} e^{2\lambda \gamma -\lambda \int_b^t \gamma_u\dee u }.
\]
As long as we can write $b$ as a monotone increasing function of $t$ such that
\[
\int_b^t \gamma_{\tau}^2 \dee \tau \to 0 \qquad \int_b^t \gamma_\tau \dee\tau \to \infty,
\]
we have the desired result that \cref{eq:simple} converges to $0$ as $t\to\infty$. 
Note that we can make $\int_b^t\gamma_\tau\dee\tau$ diverge to infinity slower by making $b$ increase faster in $t$; 
we can achieve an arbitrarily slow rate of divergence with a quickly enough increasing $b$. Therefore since $\gamma_t \to 0$ as $t\to\infty$ by assumption, we can select $b$ increasing
slowly enough that $\int_b^t\gamma_\tau\dee\tau \to \infty$, but quickly enough such that
\[
\int_b^t \gamma_\tau^2\dee \tau \leq \gamma_b \int_b^t \gamma_\tau\dee\tau \to 0.
\]
\eprfof

\bprfof{\cref{thm:converge_full}}
This result can be viewed as a special case of \cref{thm:converge_subsample} where $N=S$, and so the result can be 
obtained by following the steps of \cref{thm:converge_subsample}. Note that when $N=S$, we have that for all $t$, 
$H_t(1-s_t) = 0$. Therefore, we can obtain the desired result without \cref{assump:noisebounded} or requiring that 
$\gamma_t\to 0$ as $t\to\infty$.
\eprfof

%% file: gaussian.tex
\section{\uppercase{Gaussian location example}}\label{sec:appendix_gaussian_example}

In this section we derive an approximate bound for $\ex \kl{\pi_{w}}{\pi}$ under the Gaussian location model 
described in \cref{sec:gaussianlocation} and using the update rule in \cref{sec:alg}.
Here we use the feasible region 
\[
\fcW = \{w\in\reals^M : w\geq 0, \sum_m w_m = N\}.
\]
For better readability, we use a slightly different notation for the coreset. In particular, we use 
\[
Y = \bbmat Y_1 & \cdots & Y_M \ebmat \in \reals^{d \times M}
\] 
to denote the coreset points, 
which is a subset of the entire dataset 
\[
X = \bbmat X_1 & \cdots & X_N \ebmat \in \reals^{d \times N}.
\]
The full data posterior, as well as the coreset posterior given weights $w \in \reals^M$ can then be written as
\[
    \pi = \distNorm\lt( \frac{1}{1+N}\sum_{n=1}^N X_n,  \frac{1}{1+N}I \rt), \qquad 
    \pi_w = \distNorm\lt( \frac{\sum_{m=1}^M w_m Y_m}{1+\sum_{m=1}^M w_m}, \frac{1}{1+\sum_{m=1}^M w_m} I\rt).
\]
Since both $\pi$ and $\pi_{w_t}$ are Gaussian, we have a closed-form expression for $\kl{\pi_w}{\pi}$:
\[
\kl{\pi_w}{\pi} &= \frac{1}{2}\lt( d\log \frac{1+\sum_m w_m}{1+N} - d 
+ d\frac{1+N}{1+\sum_m w_m}
+ (1+N)\lt\|\frac{\sum_m w_m Y_m}{1+\sum_m w_m} - \frac{\sum_n X_n}{1+N}\rt\|^2\rt) \\
&= \frac{1}{2}\lt( d\log \frac{1+1^\top w}{1+N} - d 
+ d\frac{1+N}{1+1^\top w}
+ \frac{1+N}{(1+1^\top w)^2}\lt\|Yw - \frac{1+1^\top w}{1+N}X1\rt\|^2\rt)\\
&= \frac{1}{2(1+N)}\lt\|Yw - X1\rt\|^2, \label{eq:gaussiankl}
\]
where the last line is by $1^\top w = N$.
To obtain an approximate bound for $\ex \kl{\pi_{w}}{\pi}$, we look at the update rule of Coreset MCMC on $w$.
By assumption, $\beta \approx 0$, and so we assume that draws of $\theta$ obtained using the Markov kernel $\kappa_{w}$ are approximately \iid
Then following \cref{eq:gradest}, given $\theta_1, \dots, \theta_K \distiid \pi_w$, we can write the unbiased estimate of $\nabla_w \kl{\pi_w}{\pi}$ as
\[
g(w, \theta, [N]) &= \frac{-1}{K-1}\sum_{k=1}^K 
\lt( \bbmat \ell_1(\theta_k) \\ \vdots \\ \ell_M(\theta_k) \ebmat -  \bbmat \frac{1}{K}\sum_{j=1}^K\ell_1(\theta_j) \\ \vdots \\ \frac{1}{K}\sum_{j=1}^K\ell_M(\theta_j) \ebmat\rt) \cdot \\
&\quad\quad\lt(\sum_{n=1}^N \ell_n(\theta_k) - \sum_{m=1}^M w_m\ell_m(\theta_k) - \frac{1}{K}\sum_{j=1}^K\lt(\sum_{n=1}^N \ell_n(\theta_j) - \sum_{m=1}^M w_m\ell_m(\theta_j) \rt)  \rt) \\
&= \frac{-1}{K-1}\sum_{k=1}^K 
\lt( \bbmat \ell_1(\theta_k) \\ \vdots \\ \ell_M(\theta_k) \ebmat -  \bbmat \frac{1}{K}\sum_{j=1}^K\ell_1(\theta_j) \\ \vdots \\ \frac{1}{K}\sum_{j=1}^K\ell_M(\theta_j) \ebmat\rt) \cdot \\
&\quad\quad\lt( \sum_{n=1}^N \lt( \ell_n(\theta_k) - \frac{1}{K}\sum_{j=1}^K \ell_n(\theta_j) \rt) - \sum_{m=1}^M \lt( w_m\ell_m(\theta_k) - \frac{1}{K}\sum_{j=1}^K w_m\ell_m(\theta_j) \rt) \rt), \label{eq:gfull}
\]
Under the Gaussian location model, we know that
\[
\ell_n(\theta) = -\frac{1}{2}\lt\|X_n - \theta\rt\|^2.
\]
We can then simplify \cref{eq:gfull}. First note that
\[
\lt( \bbmat \ell_1(\theta_k) \\ \vdots \\ \ell_M(\theta_k) \ebmat -  \bbmat \frac{1}{K}\sum_{j=1}^K\ell_1(\theta_j) \\ \vdots \\ \frac{1}{K}\sum_{j=1}^K\ell_M(\theta_j) \ebmat\rt)
&= 
-\frac{1}{2}\lt( \bbmat \|Y_1 - \theta_k\|^2 \\ \vdots \\ \|Y_M - \theta_k\|^2 \ebmat -  \bbmat \frac{1}{K}\sum_{j=1}^K\|Y_1 - \theta_j\|^2 \\ \vdots \\ \frac{1}{K}\sum_{j=1}^K\|Y_M - \theta_j\|^2 \ebmat\rt)\\
&=
-\frac{1}{2}\lt( \bbmat -2Y_1^\top \theta_k + \|\theta_k\|^2 \\ \vdots \\ -2Y_M^\top \theta_k + \|\theta_k\|^2 \ebmat -  \bbmat \frac{1}{K}\sum_{j=1}^K-2Y_1^\top \theta_j + \|\theta_j\|^2 \\ \vdots \\ \frac{1}{K}\sum_{j=1}^K-2Y_M^\top \theta_j + \|\theta_j\|^2 \ebmat\rt)\\
&=
-\frac{1}{2}\lt( \bbmat -2Y_1^\top \lt(\theta_k - \frac{1}{K}\sum_{j=1}^K \theta_j\rt) + \|\theta_k\|^2 - \frac{1}{K}\sum_{j=1}^K \|\theta_j\|^2  \\ \vdots \\ -2Y_M^\top \lt(\theta_k - \frac{1}{K}\sum_{j=1}^K \theta_j\rt) + \|\theta_k\|^2 - \frac{1}{K}\sum_{j=1}^K \|\theta_j\|^2  \ebmat \rt)\\
&=
Y^\top \lt(\theta_k - \frac{1}{K}\sum_{j=1}^K \theta_j\rt) -\frac{1}{2} \lt(\|\theta_k\|^2 - \frac{1}{K}\sum_{j=1}^K \|\theta_j\|^2\rt) 1.
\]
We now look at the two summation terms in the last line of \cref{eq:gfull}.
\[
\sum_{n=1}^N \lt( \ell_n(\theta_k) - \frac{1}{K}\sum_{j=1}^K \ell_n(\theta_j) \rt)
&= 
-\frac{1}{2}\sum_{n=1}^N \lt( \| X_n - \theta_k\|^2 - \frac{1}{K}\sum_{j=1}^K\|X_n - \theta_j\|^2 \rt) \\
&= 
(X1)^\top \lt(\theta_k- \frac{1}{K}\sum_{j=1}^K\theta_j\rt) -\frac{N}{2}\lt(\|\theta_k\|^2 - \frac{1}{K}\sum_{j=1}^K\|\theta_j\|^2\rt).
\]
Similarly,
\[
    \sum_{m=1}^M \lt( w_m\ell_m(\theta_k) - \frac{1}{K}\sum_{j=1}^K w_m\ell_m(\theta_j) \rt)
&= 
(Yw)^\top \lt(\theta_k- \frac{1}{K}\sum_{j=1}^K\theta_j\rt) -\frac{1^\top w}{2}\lt(\|\theta_k\|^2 - \frac{1}{K}\sum_{j=1}^K\|\theta_j\|^2\rt).
\]
Therefore
\[
g(w, \theta, [N]) &= \frac{1}{K-1}\sum_{k=1}^K
\lt(Y^\top \lt(\theta_k - \frac{1}{K}\sum_{j=1}^K \theta_j\rt) -\frac{1}{2} \lt(\|\theta_k\|^2 - \frac{1}{K}\sum_{j=1}^K \|\theta_j\|^2\rt) 1\rt)\cdot\\
&\lt(\lt(\theta_k- \frac{1}{K}\sum_{j=1}^K\theta_j\rt)^\top (Yw-X1) -\frac{1^\top w-N}{2}\lt(\|\theta_k\|^2 - \frac{1}{K}\sum_{j=1}^K\|\theta_j\|^2\rt)\rt).
\]

Since $1^\top w = N$, we can use $P = \lt(I-M^{-1}11^\top \rt)$ to obtain the projected gradient update as in \cref{sec:alg}
to get
\[
w_{t+1} &= w_t - \gamma_t P g(w_t,\theta_t,[N])\\
&= w_t - 
\gamma_{t} PY^\top \frac{1}{K-1}\sum_{k=1}^K\lt(\theta_{tk} - \frac{1}{K}\sum_{j=1}^K \theta_{tj}\rt)\lt(\theta_{tk}- \frac{1}{K}\sum_{j=1}^K\theta_{tj}\rt)^\top  (Yw-X1)\\
&= w_t - \gamma_t PY^\top Q_t(Yw_t-X1),
\]
where 
\[
Q_t \defas \frac{1}{K-1}\sum_{k=1}^K\lt(\theta_{tk} - \frac{1}{K}\sum_{j=1}^K \theta_{tj}\rt)\lt(\theta_{tk}- \frac{1}{K}\sum_{j=1}^K\theta_{tj}\rt)^\top.
\]
Note that the analysis so far assumes that one does not subsample the data to estimate the gradient. 
If one were to replace the full data in $g$ with an unbiased subsample of the data, the 
projected gradient update of Coreset MCMC becomes
\[
w_{t+1} = w_t - \gamma_t PY^\top  Q_t(Yw_t-S_t), \label{eq:recursionw}
\]
where we define $S_t$ to be some random vector, independent of all else, such that $\ex S_t = X1$ and $\cov S_t = \bar\Sigma$.
By further defining $Z_t \defas Yw_t - X1$, and $A \defas YPY^\top $, the recursion in \cref{eq:recursionw}
can instead be written in terms of $Z_t$: 
\[
Z_{t+1} = (I - \gamma_t A Q_t )Z_t + \gamma_t A Q_t (S_t-X1).
\]
Solving the recursion, we have that
\[
Z_t &= \lt[\prod_{\tau=0}^{t-1} (I - \gamma_\tau A Q_\tau  )\rt] Z_0 + \sum_{\tau=0}^{t-1}\lt[\prod_{u=\tau+1}^{t-1}(I - \gamma_u A Q_u )\rt]\gamma_\tau A Q_\tau  (S_\tau-X1),
\]
where matrix products indicate left multiplication with increasing index. We can also rewrite \cref{eq:gaussiankl} as
\[
\ex \kl{\pi_{w_t}}{\pi} = \frac{1}{2(1+N)}\ex \lt\|Z_t\rt\|^2 = \frac{1}{2(1+N)}\tr\ex Z_t Z_t^\top. \label{eq:simplekl}
\]
The rest of the analysis approximates $\ex Z_t Z_t^\top$ to obtain a final approximate bound for $\ex \kl{\pi_{w_t}}{\pi}$.
With $\bm{\theta}_t$ denoting the set of $K$ samples at iteration $t$, we note that 
\[
\ex\lt[ S_t - X1 \given \bm{\theta}_1, \dots, \bm{\theta}_{t-1} \rt] = \ex\lt[ S_t - X1 \rt] = 0.
\] 
At the same time, $S_t$ is independent across all $t$ iterations.
Therefore, we can expand $\ex Z_t Z_t^\top$ to get
\[
\ex Z_tZ_t^\top  &= 
\ex \lt[\prod_{\tau=0}^{t-1} (I - \gamma_\tau A Q_\tau  )\rt]Z_0Z_0^\top \lt[\prod_{\tau=0}^{t-1} (I - \gamma_\tau A Q_\tau  )\rt]^\top \\
 &\quad+ \ex\sum_{\tau=0}^{t-1}\gamma^2_\tau \lt[\prod_{u=\tau+1}^{t-1}(I - \gamma_u A Q_u )\rt]A Q_\tau  (S_\tau-X1)(S_\tau-X1)^\top Q_\tau  ^\top A^\top \lt[\prod_{u=\tau+1}^{t-1}(I - \gamma_u A Q_u )\rt]^\top \\
&= \ex \lt[\prod_{\tau=0}^{t-1} (I - \gamma_\tau A Q_\tau  )\rt]Z_0Z_0^\top \lt[\prod_{\tau=0}^{t-1} (I - \gamma_\tau A Q_\tau  )\rt]^\top \\
 &\quad+ \sum_{\tau=0}^{t-1}\gamma^2_\tau\ex\lt[\prod_{u=\tau+1}^{t-1}(I - \gamma_u A Q_u )\rt]A Q_\tau  \bar\Sigma Q_\tau  ^\top A^\top \lt[\prod_{u=\tau+1}^{t-1}(I - \gamma_u A Q_u )\rt]^\top. \label{eq:expectations}
\]
We now use the tower property to rewrite the expectation in the second term in \cref{eq:expectations} above to get
\[
&\ex\lt[\prod_{u=\tau+1}^{t-1}(I - \gamma_u A Q_u )\rt]A Q_\tau  \bar\Sigma Q_\tau  ^\top A^\top \lt[\prod_{u=\tau+1}^{t-1}(I - \gamma_u A Q_u )\rt]^\top \\
&=
\ex\lt[\lt[\prod_{u=\tau+1}^{t-1}(I - \gamma_u A Q_u )\rt]A \ex\lt[Q_\tau  \bar\Sigma Q_\tau  ^\top \given \bm{\theta}_{\tau+1}, \dots, \bm{\theta}_{t-1} \rt] A^\top \lt[\prod_{u=\tau+1}^{t-1}(I - \gamma_u A Q_u )\rt]^\top\rt].
\]
Looking at the inner conditional expectation, we see that all randomness comes from $\theta_\tau$. 
By letting 
\[
\bar\theta_\tau = \frac{1}{K}\sum_{j=1}^K \theta_{\tau j},
\]
we then have
\[
&\ex_{\bm{\theta}_\tau} Q_\tau ^\top  \bar\Sigma Q_\tau  \\
&= \frac{1}{(1+N)^2} \ex_{\bm{\theta}_\tau} \frac{K^2}{(K-1)^2}\lt(\frac{1}{K}\sum_{k=1}^K\theta_{\tau k} \theta_{\tau k}  ^\top  -  \bar \theta_\tau   \bar \theta_\tau ^\top \rt) \bar\Sigma \lt(\frac{1}{K}\sum_{k=1}^K\theta_{\tau k} \theta_{\tau k} ^\top  -  \bar \theta_\tau   \bar \theta_\tau^\top \rt)\\
&= \frac{K^2}{(K-1)^2(1+N)^2} \ex_{\bm{\theta}_\tau} \lt(\frac{1}{K^2}\sum_{k,k'=1}^K\theta_{\tau k}  \theta_{\tau k}  ^\top \bar\Sigma \theta_{\tau k'} \theta_{\tau k'} ^\top  
- \frac{1}{K}\sum_{k=1}^K \theta_{\tau k}  \theta_{\tau k}  ^\top  \bar\Sigma \bar \theta_\tau  \bar \theta_\tau ^\top  
- \frac{1}{K}\sum_{k=1}^K \bar \theta_\tau  \bar \theta_\tau ^\top  \bar\Sigma \theta_{\tau k}  \theta_{\tau k}  ^\top  
+ \bar \theta_\tau  \bar \theta_\tau ^\top  \bar\Sigma \bar \theta_\tau  \bar \theta_\tau ^\top \rt)\\
&= \frac{K \bar\Sigma  + \tr(\bar\Sigma) I}{(K-1)(1+N)^2},
\]
where the last equality is obtained by noting that $\theta_{\tau 1}, \dots, \theta_{\tau k}$ are \iid isotropic Gaussian with variance $1/(1+N)$ and that 
\[
\ex\lt[\bar \theta_\tau\bar \theta_\tau^\top  \given \theta_{\tau j} \rt] &= \ex\lt[ \frac{1}{K^2}\sum_{k,k'=1}^K \theta_{\tau k} \theta_{\tau k'}^\top  \given \theta_{\tau j} \rt] = \frac{1}{K^2}\theta_{\tau j} \theta_{\tau j}^\top  + \frac{K-1}{K^2} I.
\]
Therefore,
\[
    &\ex\lt[\prod_{u=\tau+1}^{t-1}(I - \gamma_u A Q_u )\rt]A Q_\tau  \bar\Sigma Q_\tau  ^\top A^\top \lt[\prod_{u=\tau+1}^{t-1}(I - \gamma_u A Q_u )\rt]^\top \\
    &=
    \ex\lt[\prod_{u=\tau+1}^{t-1}(I - \gamma_u A Q_u )\rt]A \frac{K\bar\Sigma + \tr\bar\Sigma I}{(K-1)(1+N)^2}A^\top \lt[\prod_{u=\tau+1}^{t-1}(I - \gamma_u A Q_u )\rt]^\top.       
\]
Define $B \defas A \frac{K\bar\Sigma + \tr\bar\Sigma I}{(K-1)(1+N)^2}A^\top$. Note that $B$ is constant, and so we can write
\[
\ex (I - \gamma A Q_t ) B (I- \gamma A Q_t )^\top  &= \ex \lt[ B - \gamma A Q_t  B - \gamma B Q_t ^\top  A^\top  + \gamma^2 A Q_t  B Q_t ^\top  A^\top \rt] \\
&=B - \frac{\gamma}{1+N} A B - \frac{\gamma}{1+N} B A^\top   + \gamma^2 A\ex \lt[ Q_t  B Q_t ^\top\rt]A^\top \\
&=B - \frac{\gamma}{1+N} A B - \frac{\gamma}{1+N} B A^\top   + \gamma^2 A \lt(\frac{KB + \tr(B) I}{(K-1)(1+N)^2}\rt)A^\top \\
&=\lt(I-\frac{\gamma A}{1+N}\rt)B\lt(I - \frac{\gamma A}{1+N}\rt)^\top  + \gamma^2 \frac{ABA^\top  + \tr(B) AA^\top }{(K-1)(1+N)^2},
\]
where the second equality is by noting that $\ex Q_t  = \frac{1}{N+1}I$. 
Since $A$ is symmetric and positive semidefinite, we can write $A = UDU^\top $ with $D \succeq 0$ diagonal and $U$ unitary.
Then by defining $E\defas U^\top A\frac{K\bar\Sigma + \tr\bar\Sigma I}{(K-1)(1+N)^2}A^\top U$, we have
\[
B = UU^\top A\frac{K\bar\Sigma + \tr\bar\Sigma I}{(K-1)(1+N)^2}A^\top UU^\top = UEU^\top.
\]
Therefore,
\[
\ex (I - \gamma A Q_t ) B (I- \gamma A Q_t )^\top
= U\lt(\lt(I - \frac{\gamma}{1+N} D\rt) E\lt(I - \frac{\gamma}{1+N} D\rt)  + \gamma^2\frac{DED  + \tr(E) D^2}{(K-1)(1+N)^2} \rt)U^\top.
\]
From this, we can see that starting with $B = UEU^\top$, the above operation yields 
\[
\ex (I - \gamma A Q_t ) B (I- \gamma A Q_t )^\top = \ex (I - \gamma A Q_t ) UEU^\top (I- \gamma A Q_t )^\top = UE'U^\top,
\]
where each element of $E'$ can be written in terms of $E$ as 
\[
E'_{ij} = \lt\{\lt(1-\frac{\gamma}{N+1}D_i\rt)\lt(1-\frac{\gamma}{N+1}D_j\rt) + \frac{\gamma^2}{(K-1)(N+1)^2}D_iD_j\rt\}E_{ij} + \frac{\gamma^2}{(K-1)(N+1)^2} \ind[i=j] D_i^2 \tr E.
\]
Therefore, by applying this operation multiple times, we get
\[
&\ex\lt[\prod_{u=\tau+1}^{t-1}(I - \gamma_u A Q_u )\rt]A Q_\tau  \bar\Sigma Q_\tau  ^\top A^\top \lt[\prod_{u=\tau+1}^{t-1}(I - \gamma_u A Q_u )\rt]^\top \\
&=\ex\lt[\prod_{u=\tau+1}^{t-1}(I - \gamma_u A Q_u )\rt]UEU^\top \lt[\prod_{u=\tau+1}^{t-1}(I - \gamma_u A Q_u )\rt]^\top\\
&=UE''U^\top,
\]
where 
\[
E''_{ij} &= \lt[\prod_{u=\tau+1}^{t-1} \lt\{\lt(1-\frac{\gamma_u}{N+1}D_i\rt)\lt(1-\frac{\gamma_u}{N+1}D_j\rt) + \frac{\gamma_u^2}{(K-1)(N+1)^2}D_iD_j\rt\} \rt]E_{ij} \\
&\quad + \lt[\prod_{u=\tau+1}^{t-1}\frac{\gamma_u^2}{(K-1)(N+1)^2} \ind[i=j] D_i^2\rt] \tr E.
\]
Considering $\gamma_u$ small enough such that $\gamma^2_u/(N+1)^2 \ll \gamma_u / (N+1)$, we drop the higher order terms to find that
\[
&\ex\lt[\prod_{u=\tau+1}^{t-1}(I - \gamma_u A Q_u )\rt]A Q_\tau  \bar\Sigma Q_\tau  ^\top A^\top \lt[\prod_{u=\tau+1}^{t-1}(I - \gamma_u A Q_u )\rt]^\top\\
&\approx U \lt[\prod_{u=\tau+1}^{t-1}\lt(I - \frac{\gamma_u}{N+1}D\rt)\rt] U^\top A\frac{K\bar\Sigma + \tr\bar\Sigma I}{(K-1)(1+N)^2}A^\top U \lt[\prod_{u=\tau+1}^{t-1}\lt(I - \frac{\gamma_u}{N+1}D\rt)\rt]^\top U^\top \\
&\approx e^{-\frac{\sum_{u=\tau+1}^{t-1} \gamma_u}{N+1}A} A \frac{K\bar\Sigma + \tr\bar\Sigma I}{(K-1)(1+N)^2}A^\top e^{- \frac{\sum_{u=\tau+1}^{t-1}\gamma_u}{N+1}A^\top }.
\]
For the expectation in the first term in \cref{eq:expectations}, since $Z_0Z_0^\top$ is constant, 
we can use the same trick to get
\[
\ex \lt[\prod_{\tau=0}^{t-1} (I - \gamma_\tau A Q_\tau  )\rt]Z_0Z_0^\top \lt[\prod_{\tau=0}^{t-1} (I - \gamma_\tau A Q_\tau  )\rt]^\top
\approx e^{-\frac{\sum_{\tau=0}^{t-1}\gamma_\tau}{N+1}A}Z_0Z_0^\top  e^{-\frac{\sum_{\tau=0}^{t-1}\gamma_\tau}{N+1}A^\top }.
\]
Therefore,
\[
\ex Z_tZ_t^\top  &\approx
e^{-\frac{\sum_{\tau=0}^{t-1}\gamma_\tau}{N+1}A}Z_0Z_0^\top  e^{-\frac{\sum_{\tau=0}^{t-1}\gamma_\tau}{N+1}A^\top }+ 
\sum_{\tau=0}^{t-1}\frac{\gamma^2_\tau e^{-\frac{\sum_{u=\tau+1}^{t-1}\gamma_u}{N+1}A}A (K\bar\Sigma + \tr\bar\Sigma I)A^\top e^{-\frac{\sum_{u=\tau+1}^{t-1}\gamma_u}{N+1}A^\top }}{(K-1)(N+1)^2}.
\]

Note that for uniform independent subsamples of size $S$ 
without replacement,
\[
\bar\Sigma = \frac{N^2(N-S)}{S(N-1)}\lt[ \frac{1}{N}XX^\top  - \frac{1}{N^2}X11^\top X^\top \rt] = \frac{N^2(N-S)}{S(N-1)}\Sigma_X,
\]
where $\Sigma_X = \frac{1}{N}XX^\top  - \frac{1}{N^2}X11^\top X^\top  = \frac{1}{N}X\lt(I-\frac{1}{N}11^\top \rt)X^\top $ is the empirical covariance of the full data.
Similarly we have $A = M \Sigma_Y$ for the empirical covariance of the coreset.
Therefore, 
\[
\ex Z_tZ_t^\top  &\approx
e^{-\frac{M\sum_{\tau=0}^{t-1}\gamma_\tau}{N+1}\Sigma_Y}Z_0Z_0^\top  e^{-\frac{M\sum_{\tau=0}^{s-1}\gamma_\tau}{N+1}\Sigma_Y^\top }+ \\
&\quad\sum_{\tau=0}^{s-1}\frac{\gamma^2_\tau M^2 N^2(N-S)e^{-\frac{M\sum_{u=\tau+1}^{t-1}\gamma_u}{N+1}\Sigma_Y}\Sigma_Y (K\Sigma_X + \tr\Sigma_X I)\Sigma_Y^\top e^{-\frac{M\sum_{u=\tau+1}^{t-1}\gamma_u}{N+1}\Sigma_Y^\top }}{(N-1)S(K-1)(N+1)^2}.
\]
By assumption, $N \gg M \gg 1$. Since the data are generated \iid from a standard Gaussian,
we can approximate $\Sigma_X \approx \Sigma_Y \approx I$.
Therefore, 
\[
\ex Z_tZ_t^\top  &\approx
e^{-\frac{M\lt(\sum_{\tau=0}^{t-1}\gamma_\tau+\sum_{\tau=0}^{t-1}\gamma_\tau\rt)}{N+1}}Z_0Z_0^\top  
+ \frac{M^2 (N-S)N^2(K+d)}{(N-1)S(K-1)(N+1)^2}\sum_{\tau=0}^{t-1}\gamma^2_\tau e^{-\frac{M\lt(\sum_{u=\tau+1}^{t-1}\gamma_u+\sum_{u=\tau+1}^{t-1}\gamma_u\rt)}{N+1}} I.
\] 
We assume that $\gamma_t = \gamma (t+1)^{\alpha-1}$ for some $\gamma>0$ and $0 \leq \alpha \leq 1$. Then we have the approximation
\[
\sum_{\tau=0}^{t-1}\gamma_\tau \approx \gamma \int_0^{t-1}(\tau+1)^{\alpha-1}\dee\tau = \frac{\gamma (t^\alpha - 1)}{\alpha}.
\]
Using this approximation and the fact that $N+1 \approx N-1 \approx N$, we obtain
\[
\ex Z_tZ_t^\top  &\approx
e^{-\frac{\gamma M\lt(t^\alpha + t^\alpha-2\rt)}{\alpha N}}Z_0Z_0^\top  + \frac{(N-S)M^2(K+d)}{NS(K-1)}\sum_{\tau=1}^{t} \gamma_{\tau-1}^2 e^{-\frac{\gamma M\lt(t^\alpha + t^\alpha - 2\tau^\alpha\rt)}{\alpha N}} I\\
&= e^{-\frac{\gamma M\lt(t^\alpha + t^\alpha-2\rt)}{\alpha N}}Z_0Z_0^\top  + \frac{\gamma^2(N-S)M^2(K+d)}{NS(K-1)}\sum_{\tau=1}^{t}\tau^{2(\alpha-1)} e^{-\frac{\gamma M\lt(t^\alpha + t^\alpha - 2\tau^\alpha\rt)}{\alpha N}} I.
\]
Again using an approximation of the sum as an integral, we get
\[
\ex Z_tZ_t^\top  &\approx
e^{-\frac{\gamma M\lt(t^\alpha + t^\alpha-2\rt)}{\alpha N}}Z_0Z_0^\top  + \frac{\gamma^2(N-S)M^2(K+d)}{NS(K-1)}\int_1^t \tau^{2(\alpha-1)} e^{-\frac{\gamma M\lt(t^\alpha + t^\alpha - 2\tau^\alpha\rt)}{\alpha N}} \dee\tau I.
\]
We can now substitute the above expression to \cref{eq:simplekl} to get
\[
\ex \kl{\pi_{w_t}}{\pi}
&=
\frac{1}{2(N+1)} \tr \ex Z_tZ_t^\top\\
&\approx \frac{1}{2(N+1)} \ex \|Z_0\|^2 + \frac{\gamma^2(N-S)M^2(K+d)}{NS(K-1)}\int_1^t \tau^{2(\alpha-1)} e^{-\frac{\gamma M\lt(t^\alpha + t^\alpha - 2\tau^\alpha\rt)}{\alpha N}} \dee\tau \tr I\\
&\approx \frac{1}{2(N+1)} \ex \|Z_0\|^2 + \frac{\gamma^2(N-S)M^2(K+d)d}{NS(K-1)}\int_1^t \tau^{2(\alpha-1)} e^{-\frac{\gamma M\lt(t^\alpha + t^\alpha - 2\tau^\alpha\rt)}{\alpha N}} \dee\tau.
\]
Again by the assumption that $N \gg M \gg 1$ and that all data points are \iid standard Gaussian, we have $\ex\|Z_0\|^2 \approx N^2/M$.
Therefore,
\[
\ex \kl{\pi_{w_t}}{\pi} \approx e^{-\frac{2\gamma M\lt(t^\alpha - 1\rt)}{\alpha N }}\frac{N}{2M} + \frac{\gamma^2(N-S)M^2(K+d)d}{2N^2S(K-1)}\int_1^t \tau^{2(\alpha-1)} e^{-\frac{2\gamma M\lt(t^\alpha - \tau^\alpha\rt)}{\alpha N }} \dee\tau.
\]
Set $\gamma = c N/M$ for some tunable $0 < c \ll M$, then 
\[
\ex \kl{\pi_{w_t}}{\pi} 
&\approx 
e^{-\frac{2c\lt(t^\alpha - 1\rt)}{\alpha }}\frac{N}{2M} + \frac{c^2(N-S)(K+d)d}{2S(K-1)}\int_1^t \tau^{2(\alpha-1)} e^{-\frac{2c\lt(t^\alpha - \tau^\alpha\rt)}{\alpha }} \dee\tau. \label{eq:integral}
\]
We now derive an approximate upper bound for the integral term in \cref{eq:integral} above.
Apply the transformation of variables $x = \tau^\alpha$ with
$\dee x = \alpha \tau^{\alpha-1} \dee \tau$ to get
\[
\int_1^t \tau^{2(\alpha-1)} e^{-\frac{2c\lt(t^\alpha - \tau^\alpha\rt)}{\alpha }} \dee\tau 
&= e^{-\frac{2ct^\alpha}{\alpha }}\int_1^t \tau^{2(\alpha-1)} e^{\frac{2c\tau^\alpha}{\alpha }} \dee\tau\\
&= \alpha^{-1}e^{-\frac{2ct^\alpha}{\alpha }}\int_1^{t^\alpha} x^{1-\alpha^{-1}} e^{\frac{2cx}{\alpha }} \dee x.
\]
Let $1 \leq b \leq t$ and $b \to \infty$ as $t\to\infty$, and so
\[
 \int_1^{b^\alpha} x^{1-\alpha^{-1}} e^{\frac{2cx}{\alpha }} \dee x \leq \int_1^{b^\alpha} e^{\frac{2cx}{\alpha }} \dee x \leq \frac{\alpha}{2c}\lt(e^{\frac{2cb^\alpha}{\alpha}} - e^{\frac{2c}{\alpha}}\rt) \leq \frac{\alpha}{2c}e^{\frac{2cb^\alpha}{\alpha}}.
\]
We also have that
\[
\int_{b^\alpha}^{t^\alpha} x^{1-\alpha^{-1}} e^{\frac{2cx}{\alpha }} \dee x \leq e^{\frac{2ct^\alpha}{\alpha }} \int_{b^\alpha}^{t^\alpha} x^{1-\alpha^{-1}} \dee x = e^{\frac{2ct^\alpha}{\alpha }}\frac{t^{2\alpha-1} - b^{2\alpha-1}}{2-\alpha^{-1}}.
\]
Therefore for any $1 \leq b \leq t$, 
\[
\int_1^t \tau^{2(\alpha-1)} e^{-\frac{2c(t^\alpha-\tau^\alpha)}{\alpha }} \dee\tau
&\leq \alpha^{-1}e^{-\frac{2ct^\alpha}{\alpha }}\lt( \frac{\alpha}{2c}e^{\frac{2cb^\alpha}{\alpha}} + 
e^{\frac{2ct^\alpha}{\alpha }}\frac{t^{2\alpha-1} - b^{2\alpha-1}}{2-\alpha^{-1}} \rt)\\
&= \frac{1}{2c}e^{-\frac{2c(t^\alpha-b^\alpha)}{\alpha}} + \frac{t^{2\alpha-1} - b^{2\alpha-1}}{2\alpha-1}.
\]
Then let $b = t - h(t)$ where $h(t) = o(t)$. So as $t$ gets large, $b \approx t$, and so
\[
\frac{1}{2c}e^{-\frac{2c(t^\alpha-b^\alpha)}{\alpha}} + \frac{t^{2\alpha-1} - b^{2\alpha-1}}{2\alpha-1}
&\approx 
\frac{1}{2c}e^{-\frac{2ct^{\alpha-1}h(t)}{\alpha}} + \frac{t^{2\alpha-2}h(t)}{2\alpha-1}\\
&\leq \frac{1}{2c}e^{-2ct^{\alpha-1}h(t)} + t^{2\alpha-2}h(t).
\]
By setting $h(t) = \frac{1-\alpha}{2c}t^{1-\alpha}\log t$,
\[
\int_1^t \tau^{2(\alpha-1)} e^{-\frac{2c(t^\alpha-\tau^\alpha)}{\alpha }} \dee\tau
&\lessapprox
\frac{1}{2ct^{1-\alpha}} + \frac{1-\alpha}{2ct^{1-\alpha}}\log t = \frac{1+(1-\alpha)\log t}{2ct^{1-\alpha}}.
\]
By substituting the above approximate bound to \cref{eq:integral}, we arrive at the final approximate bound
\[
\ex \kl{\pi_{w_t}}{\pi} 
&\lessapprox
e^{-\frac{2c\lt(t^\alpha - 1\rt)}{\alpha }}\frac{N}{2M} + \frac{c(N-S)(K+d)d}{4S(K-1)}\frac{(1+(1-\alpha)\log t)}{t^{1-\alpha}}.
\]

%% file: experimental_details.tex
\section{\uppercase{Details of experiments}}
\label{sec:expt}

\subsection{Model Specification}
In this subsection, we describe the four examples (three real data and one synthetic) that we used for our experiments. 
For each model, we are given a set of points $(x_n,y_n)^N_{n=1}$, each consisting of features $x_n\in\reals^p$ and 
response $y_n$.

\textbf{Bayesian linear regression:} We use the model
\[
  &\begin{bmatrix} \beta & \log\sigma^2 \end{bmatrix}^\top \dist \distNorm(0,I),\\ 
  &\forall n\in[N],
  y_n \given x_n, \beta, \sigma^2 \distind \distNorm\left(\begin{bmatrix} 1 & x_n^\top \end{bmatrix}\beta, \sigma^2\right),
\]
where $\beta\in\reals^{p+1}$ is a vector of regression coefficients and $\sigma^2\in\reals_+$ is the noise variance. 
Note that the prior here is not conjugate for the likelihood.
The dataset\footnote{The dataset is available at 
\url{https://github.com/NaitongChen/Sparse-Hamiltonian-Flows}.}
consists of flight delay information from $N=$ 98,673 observations. We study 
the difference, in minutes, between the scheduled and actual departure times against $p=10$ features 
including flight-specific and meteorological information.

\textbf{Bayesian logistic regression:} We use the model
\[
    &\forall i\in[p+1], \quad \beta_i \distiid \distCauchy(0,1), \\
	&\forall n\in[N], \quad y_n \distind \distBern 
    \left(\left(1+\exp\left(-\begin{bmatrix} 1 & x_n^\top\end{bmatrix} \beta\right)\right)^{-1}\right),  
\]
where $\beta = \begin{bmatrix}\beta_1 & \dots & \beta_{p+1} \end{bmatrix}^\top\in\reals^{p+1}$ is a vector of regression coefficients. Here we use the same dataset as in linear 
regression, but instead model the relationship between whether a flight is cancelled using the same set of features. 
Note that of all flights included, only $0.058\%$ were cancelled.

\textbf{Bayesian Poisson regression:} We use the model
\[
  &\beta \dist \distNorm(0, I),\\
  &\forall n\in[N],
  y_n \given x_n, \beta \distind 
  \distPoiss\left( \log\left( 1 + e^{ \begin{bmatrix} 1 & x_n^\top \end{bmatrix}\beta } \right) \right),
\]
where $\beta\in\reals^{p+1}$ is a vector of regression coefficients. Here we use a processed version of the 
bikeshare dataset\footnote{The dataset is available at 
\url{https://github.com/trevorcampbell/bayesian-coresets}.}
consisting of $N=$ 15,641 
data points. We model the hourly count of rental bikes against $p=8$ features (e.g., temperature, humidity at the 
time, and whether or not the day is a workday).

\textbf{Bayesian sparse linear regression:} This is based on Example 4.1 from \cite{george1993variable}. We use the model
\[
  &\sigma^2 \dist \distInvGam\left(\nu/2, \nu\lambda/2\right),\\
  \forall i\in[p], \quad &\gamma_i \distiid \distBern(q), \\
  &\beta_i \given \gamma_i \distind \distNorm\left(0, \left(\ind(\gamma_i = 0)\tau + \ind(\gamma_i = 1)c\tau\right)^2 \right), \\ 
  \forall n\in[N], \quad &y_n \given x_n, \beta, \sigma^2 \distind \distNorm\left( x_n^\top\beta, \sigma^2 \right),
\]
where we set $\nu=0.1, \lambda=1, q=0.1, \tau=0.1$, and $c=10$. Here we model the variance $\sigma^2$, the vector of regression 
coefficients $\beta = \begin{bmatrix}\beta_1 & \dots & \beta_p \end{bmatrix}^\top\in\reals^p$ and  
the vector of binary variables $\gamma = \begin{bmatrix}\gamma_1 & \dots & \gamma_p \end{bmatrix}^\top\in\left\{0,1\right\}^p$ 
indicating the inclusion of the $p^\text{th}$ feature in the model.
We set $N=50{,}000$, $p=10$, $\beta^\star = \begin{bmatrix}0 & 0 & 0 & 0 & 0 & 5 & 5 & 5 & 5 & 5\end{bmatrix}^\top$, 
and generate a synthetic dataset by
\[
  \forall n \in [N], \quad &x_n \distiid \distNorm\left( 0, I \right),\\
  &\eps_n \distiid \distNorm\left( 0, 25^2 \right),\\
  &y_n = x_n^\top \beta^\star + \eps_n.
\]

For full-data inference results of the three real data examples, we ran Stan \citep{carpenter2017stan} with 10 parallel chains,
each taking $15{,}000$ steps where the first $5{,}000$ were discarded, for a combined $100{,}000$ draws. 
Full-data inference time was $\sim 22$ minutes for linear regression, $\sim 28$ minutes for logistic regression, 
and $\sim 5$ minutes for Poisson regression,
though note that these times are not comparable with the other results since Stan is highly optimized. 
For full-data inference result of the synthetic data example, we use the Gibbs sampler developed by \cite{george1993variable}
to generate $200{,}000$ draws, with the first half discarded as burn-in. Full-data inference time for the sparse linear regression 
model was $\sim 20$ minutes.

\subsection{Parameter Settings}
We begin by describing settings that apply across all experiments before moving to model-specific settings.
For each method, we use $10,000$ samples to estimate metrics that assess posterior approximation quality and sampling 
efficiency. For metrics that require information from the full-data posterior, we use a sample of size $100,000$ from Stan 
\citep{carpenter2017stan} across all three real data experiments, and a sample of size $10,000$ using the Gibbs sampler 
developed by \cite{george1993variable} for the synthetic data experiment.
To account for changes in $w$, for all coreset methods that use an MCMC kernel, we use the hit-and-run slice sampler 
with doubling \citep{belisle1993hit,neal2003slice} for linear and logistic regressions, the univariate slice sampler 
with doubling (\citealp[Fig.~4]{neal2003slice}) applied to each dimension for the mor challenging Poisson regression, 
and the Gibbs sampler developed by \cite{george1993variable} for sparse regression. 
Note that to get the $10,000$ samples for metric evaluation from each method, we simulate $20,000$ MCMC states and 
take the second half of the 
chain to ensure proper mixing. The exceptions are \texttt{CoresetMCMC} and \texttt{SHF}. For \texttt{CoresetMCMC}, 
no burn-in is necessary as in each of our experiments we adapt the coreset weights for more than $10,000$ steps. 
For \texttt{SHF}, we have access to \iid samples from the final trained posterior approximation.

Across all experiments, we set $K=2$ for \texttt{CoresetMCMC}. We follow \cite{chen2022bayesian} and use $8$ 
quasi-refreshment and $10$ leapfrog steps between quasi-refreshments in \texttt{SHF}. To estimate the ELBO objective, 
we use a fresh minibatch of size $100$ at each optimization iteration. Both \texttt{CoresetMCMC} and 
\texttt{SHF} are trained using ADAM \citep{kingma2014adam}. For \texttt{QNC}, we use a sample of size $1,000$ to 
estimate each weight update. We set the threshold to $0.05$ for both \texttt{Austerity} and \texttt{Confidence}. For 
tuning \texttt{SGLD-CV} and \texttt{SGHMC-CV}, we follow \cite{baker2019sgmcmc}, and find the control variate 
using ADAM.

\textbf{Bayesian linear regression:}
We train \texttt{CoresetMCMC} for $25{,}000$ iterations with the ADAM step sizes set to $20$, $1$, $10$, $10$, $1$, $1$ for 
coreset sizes $M=10$, $20$, $50$, $100$, $200$, $500$. 
We also train \texttt{CoresetMCMC-S} for $25{,}000$ iterations with the ADAM step sizes set to 
$10t^{-0.1}$, $10t^{-0.1}$, $10t^{-0.1}$, $10t^{-0.3}$, $20t^{-0.5}$, $20t^{-0.3}$ for 
coreset sizes $M=10$, $20$, $50$, $100$, $200$, $500$, 
where $t$ is the iteration number. At each gradient estimate, we set the minibatch size to be $5$ times the coreset 
size.
We train \texttt{QNC} for $50$ iterations so that the total number of samples drawn from the coreset posterior during 
optimization is the same between \texttt{CoresetMCMC} and \texttt{QNC}. 
In \texttt{QNC}, we choose the step size using a line search for each 
of the first $10$ iterations. For the line search procedure, we use the curvature part of the Wolfe
condition \citep{wolfe1969convergence} to shrink the step size for a maximum of $20$ times. For all iterations after, 
we use the same step size that is used in th $10^\text{th}$ iteration. For \texttt{SHF}, we set the Adam step size 
to $0.002$ and train the flow for $50,000$ iterations. We use standard normal as the initial distribution across all 
dimensions.

For \texttt{Austerity}, we set the minibatch size to be $100$. For both \texttt{Austerity} and \texttt{Confidence}, we 
use independent Gaussian proposal centred at the previous state with the variance of each component set to $0.01$. For 
both \texttt{SGLD-CV} and \texttt{SGHMC-CV}, we set the subsample size to $500$. 
For \texttt{SGLD-CV}, we use a constant step size of $2\times 10^{-5}$
For \texttt{SGHMC-CV}, we use $30$ 
leapfrog steps and a constant step size of $8\times 10^{-4}$. 
To find the mode for the control variate, we use ADAM with default settings, drawing a subsamples of $100$ observations 
until the weight updates for all dimensions are of magnitude smaller than $5\times 10^{-5}$.
All other parameters for the two stochastic gradient MCMC 
methods follow that of \cite{baker2019sgmcmc}. 

\textbf{Bayesian logistic regression:}
We train \texttt{CoresetMCMC} for $25{,}000$ iterations with the ADAM step sizes set to $0.1$, $5$, $1$, $1$, $1$, $0.1$ for 
coreset sizes $M=10$, $20$, $50$, $100$, $200$, $500$. 
We also train \texttt{CoresetMCMC-S} for $25{,}000$ iterations with the ADAM step sizes set to 
$10t^{-0.1}$, $10t^{-0.1}$, $10t^{-0.1}$, $10t^{-0.3}$, $t^{-0.3}$, $10t^{-0.5}$ for 
coreset sizes $M=10$, $20$, $50$, $100$, $200$, $500$,
where $t$ is the iteration number. 
For both \texttt{Austerity} and \texttt{Confidence}, we 
use independent Gaussian proposal centred at the previous state with the variance of each component set to $0.001$.  
To find the mode for the control variate for both stochastic gradient MCMC methods, we use ADAM with default settings, 
drawing a subsamples of $100$ observations 
until the weight updates for all dimensions are of magnitude smaller than $1\times 10^{-3}$.
All other parameters are the same as in Bayesian linear regression.

To account for the class imbalance problem, for all coreset methods, we include all observations from the rare positive 
class if the coreset size is more than twice as big as the total number of observations with positive labels. 
Otherwise we sample our coreset to have $50\%$ positive labels and $50\%$ negative labels. This is only done for 
coreset methods because unlike subsampling MCMC methods which gets a fresh subsample each time, coreset points are 
only picked once.

\textbf{Bayesian Poisson regression:}
We train \texttt{CoresetMCMC} for $50{,}000$ iterations with the ADAM step sizes set to $2$, $0.5$, $0.5$, $0.1$, $0.05$, $0.01$ for 
coreset sizes $M=10$, $20$, $50$, $100$, $200$, $500$. 
We also train \texttt{CoresetMCMC-S} for $50{,}000$ iterations with the ADAM step sizes set to 
$t^{-0.1}$, $2t^{-0.1}$, $2t^{-0.1}$, $t^{-0.1}$, $2t^{-0.3}$, $0.5t^{-0.3}$ for 
coreset sizes $M=10$, $20$, $50$, $100$, $200$, $500$, 
where $t$ is the iteration number. 
At each gradient estimate, we set the minibatch size to be $10$ times the coreset size.
We train \texttt{QNC} for $100$ iterations so that the total number of samples drawn from the coreset posterior during 
optimization is the same between \texttt{CoresetMCMC} and \texttt{QNC}. 
In \texttt{QNC}, we choose the step size using a line search for each 
of the first $10$ iterations, where the step size may be shrunk up to $50$ times each.
For both \texttt{Austerity} and \texttt{Confidence}, we 
use independent Gaussian proposal centred at the previous state with the variance of each component set to $0.01$. 
For \texttt{SGLD-CV}, we use a constant step size of $2\times 10^{-4}$
For \texttt{SGHMC-CV}, we use $30$ 
leapfrog steps and a constant step size of $8\times 10^{-4}$. 
To find the mode for the control variate, we use ADAM with default settings, drawing a subsamples of $1000$ observations 
until the weight updates for all dimensions are of magnitude smaller than $1\times 10^{-4}$.
All other parameters are the same as in Bayesian linear regression.

\textbf{Bayesian sparse linear regression:}
We train \texttt{CoresetMCMC} for $25{,}000$ iterations with the ADAM step sizes set to $0.1$, $0.1$, $1$, $1$, $0.1$, $0.01$ for 
coreset sizes $M=10$, $20$, $50$, $100$, $200$, $500$. 
We also train \texttt{CoresetMCMC-S} for $25{,}000$ iterations with the ADAM step sizes set to 
$5t^{-0.1}$, $10t^{-0.1}$, $10t^{-0.1}$, $t^{-0.1}$, $5t^{-0.5}$, $2t^{-0.5}$ for 
coreset sizes $M=10$, $20$, $50$, $100$, $200$, $500$,
where $t$ is the iteration number. 
At each gradient estimate, we set the minibatch size to be $10$ times the coreset size.
For the continuous components of both \texttt{Austerity} and \texttt{Confidence}, we 
use independent truncated Gaussian proposal that are centred at the previous state with the variance of each component set to $0.001$. 
For the discrete components, we use independent Bernoulli proposals with the success probability set to $0.5$. 
All other parameters are the same as in Bayesian linear regression.
Note that due to the inclusion of discrete variables, 
\texttt{SHF}, \texttt{SGLD-CV}, and \texttt{SGHMC-CV} are no longer directly applicable, and are hence excluded 
from this experiment.

%% file: additional_results.tex
\captionsetup[subfigure]{labelformat=empty}
\begin{figure*}[t!]
\centering
\begin{subfigure}[b]{0.24\textwidth}
    \scalebox{1}{\includegraphics[width=\textwidth]{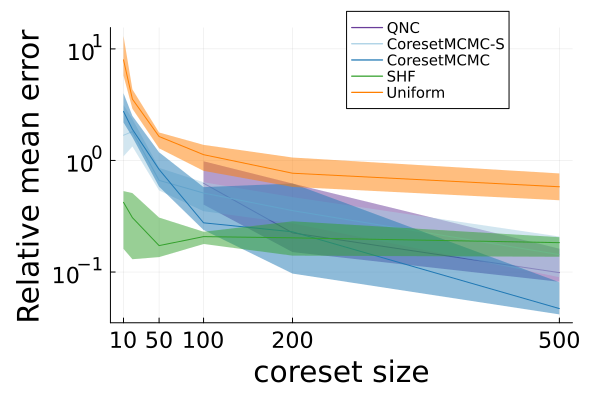}}
    \caption{(a) linear regression \label{fig:lin_mrel_coreset}}
\end{subfigure}
\hfill
\centering
\begin{subfigure}[b]{0.24\textwidth}
    \scalebox{1}{\includegraphics[width=\textwidth]{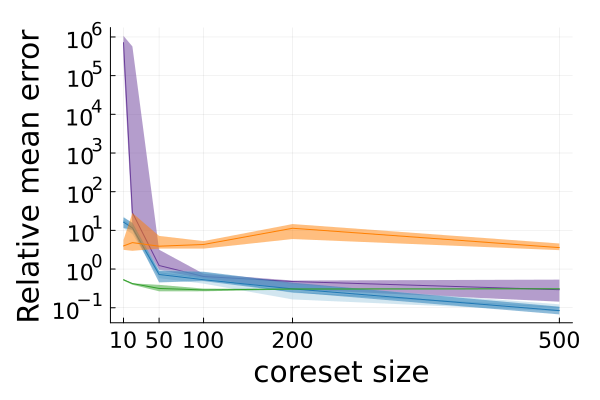}}
    \caption{(b) logistic regression \label{fig:log_mrel_coreset}}
\end{subfigure}
\hfill
\centering
\begin{subfigure}[b]{0.24\textwidth}
    \scalebox{1}{\includegraphics[width=\textwidth]{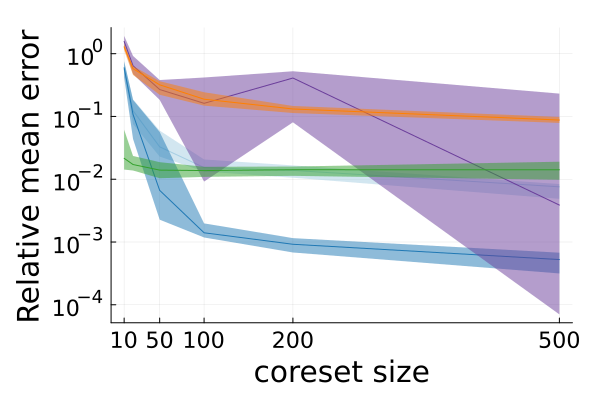}}
    \caption{(c) Poisson regression \label{fig:poi_mrel_coreset}}
\end{subfigure}
\hfill
\centering
\begin{subfigure}[b]{0.24\textwidth}
    \scalebox{1}{\includegraphics[width=\textwidth]{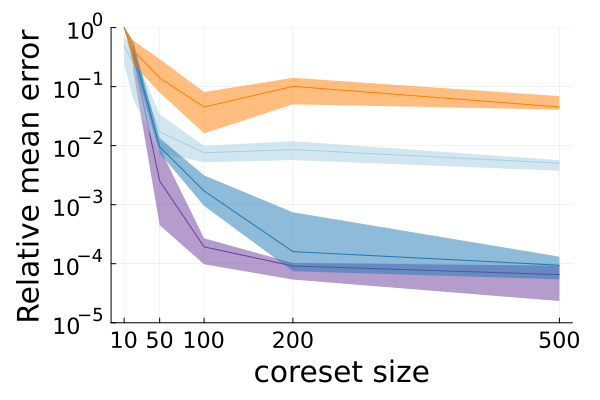}}
    \caption{(d) sparse regression \label{fig:sp_mrel_coreset}}
\end{subfigure}
\hfill
\centering
\begin{subfigure}[b]{0.24\textwidth}
    \scalebox{1}{\includegraphics[width=\textwidth]{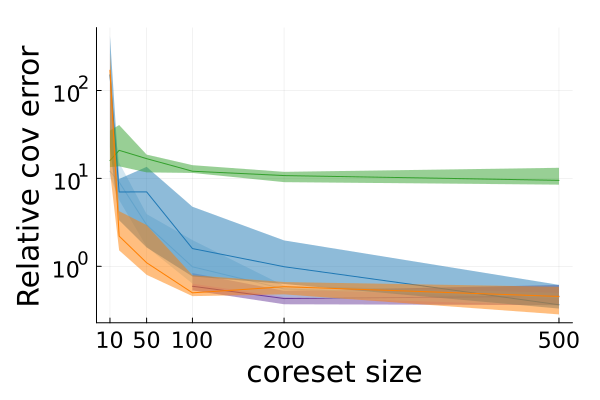}}
    \caption{(e) linear regression \label{fig:lin_srel_coreset}}
\end{subfigure}
\hfill
\centering
\begin{subfigure}[b]{0.24\textwidth}
    \scalebox{1}{\includegraphics[width=\textwidth]{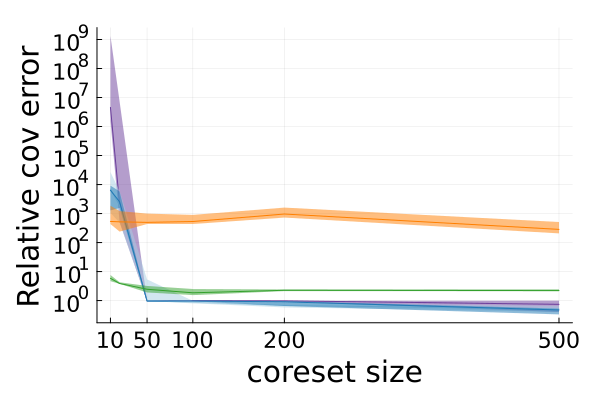}}
    \caption{(f) logistic regression \label{fig:log_srel_coreset}}
\end{subfigure}
\hfill
\centering
\begin{subfigure}[b]{0.24\textwidth}
    \scalebox{1}{\includegraphics[width=\textwidth]{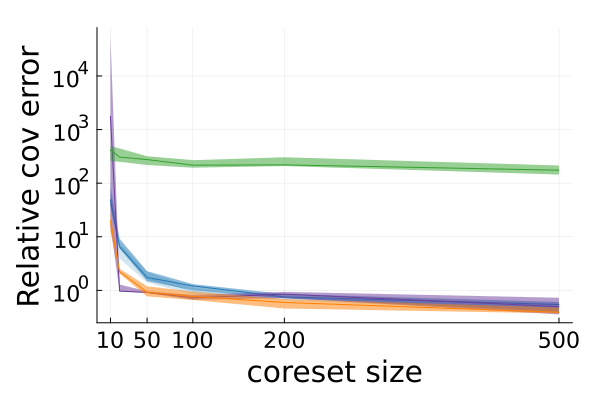}}
    \caption{(g) Poisson regression \label{fig:poi_srel_coreset}}
\end{subfigure}
\centering
\begin{subfigure}[b]{0.24\textwidth}
    \scalebox{1}{\includegraphics[width=\textwidth]{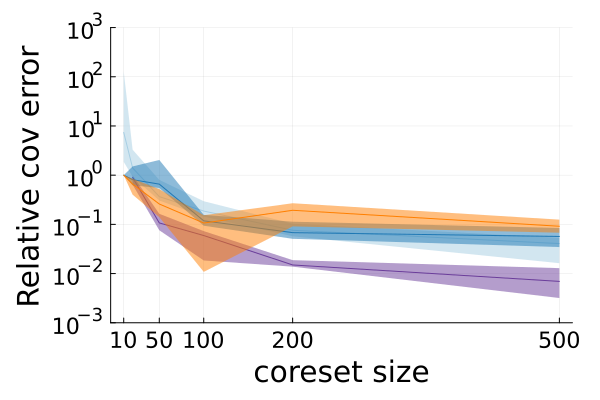}}
    \caption{(h) sparse regression \label{fig:sp_srel_coreset}}
\end{subfigure}

\caption{
Comparison of coreset methods. \cref{fig:lin_mrel_coreset,fig:log_mrel_coreset,fig:poi_mrel_coreset,fig:sp_mrel_coreset} show
posterior approximation quality via the relative mean error, and \cref{fig:lin_srel_coreset,fig:log_srel_coreset,fig:poi_srel_coreset,fig:sp_srel_coreset}
show posterior approximation quality via the relative covariance error. The lines indicate the median, and error regions indicate 25$^\text{th}$ to 75$^\text{th}$ percentile from 10 runs.
}
\label{fig:msrel_coreset}
\end{figure*}

\captionsetup[subfigure]{labelformat=empty}
\begin{figure*}[t!]
\centering
\begin{subfigure}[b]{0.24\textwidth}
    \scalebox{1}{\includegraphics[width=\textwidth]{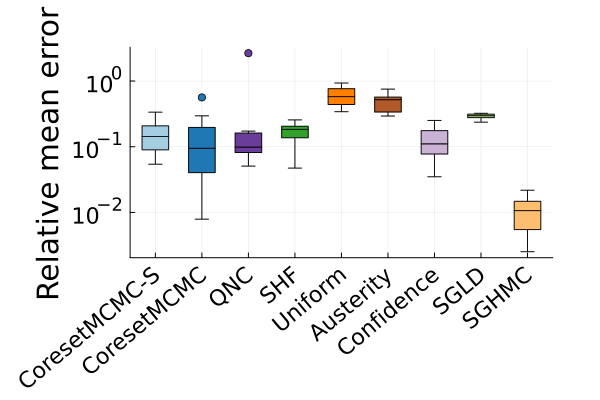}}
    \caption{(a) linear regression \label{fig:lin_mrel_all}}
\end{subfigure}
\hfill
\centering
\begin{subfigure}[b]{0.24\textwidth}
    \scalebox{1}{\includegraphics[width=\textwidth]{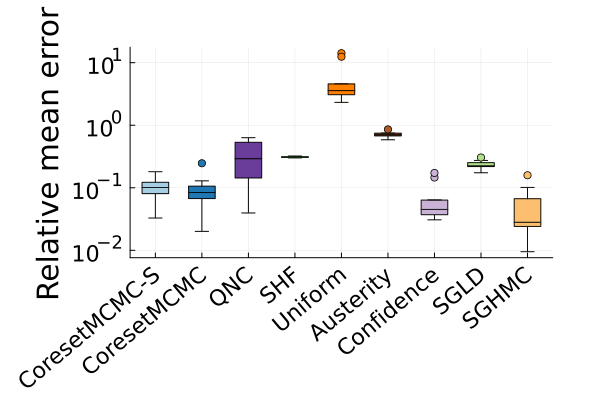}}
    \caption{(b) logistic regression \label{fig:log_mrel_all}}
\end{subfigure}
\hfill
\centering
\begin{subfigure}[b]{0.24\textwidth}
    \scalebox{1}{\includegraphics[width=\textwidth]{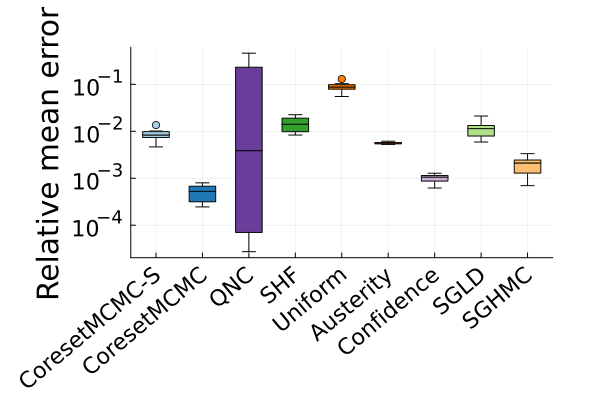}}
    \caption{(c) Poisson regression \label{fig:poi_mrel_all}}
\end{subfigure}
\hfill
\centering
\begin{subfigure}[b]{0.24\textwidth}
    \scalebox{1}{\includegraphics[width=\textwidth]{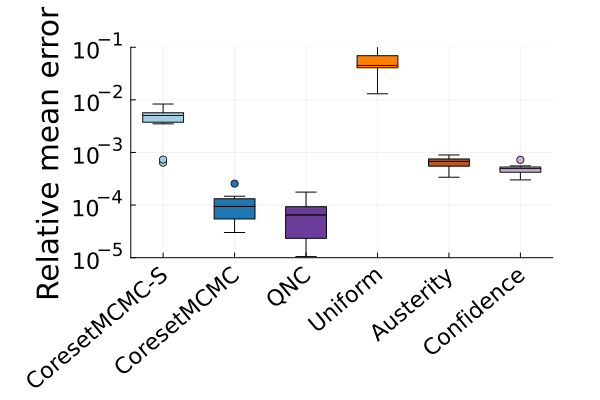}}
    \caption{(d) sparse regression \label{fig:sp_mrel_all}}
\end{subfigure}
\hfill
\centering
\begin{subfigure}[b]{0.24\textwidth}
    \scalebox{1}{\includegraphics[width=\textwidth]{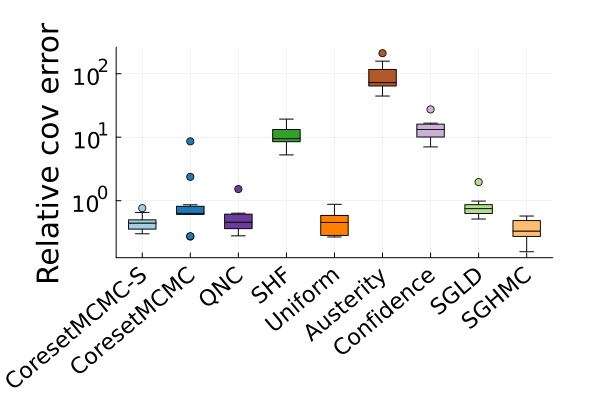}}
    \caption{(e) linear regression \label{fig:lin_srel_all}}
\end{subfigure}
\hfill
\centering
\begin{subfigure}[b]{0.24\textwidth}
    \scalebox{1}{\includegraphics[width=\textwidth]{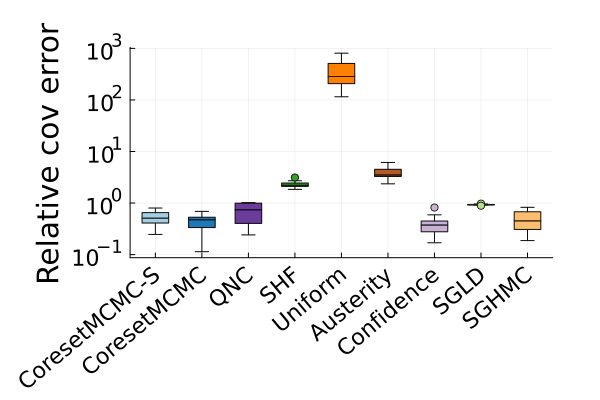}}
    \caption{(f) logistic regression \label{fig:log_srel_all}}
\end{subfigure}
\hfill
\centering
\begin{subfigure}[b]{0.24\textwidth}
    \scalebox{1}{\includegraphics[width=\textwidth]{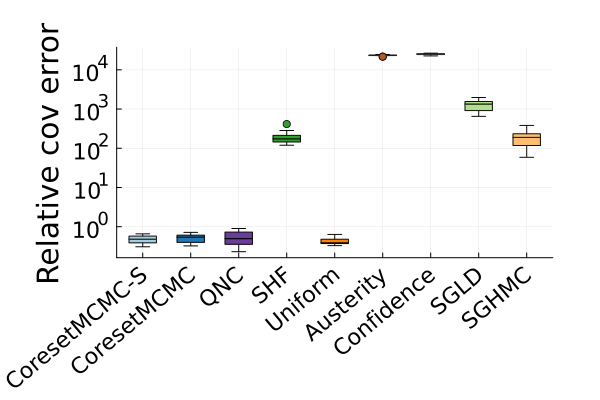}}
    \caption{(g) Poisson regression \label{fig:poi_srel_all}}
\end{subfigure}
\hfill
\centering
\begin{subfigure}[b]{0.24\textwidth}
    \scalebox{1}{\includegraphics[width=\textwidth]{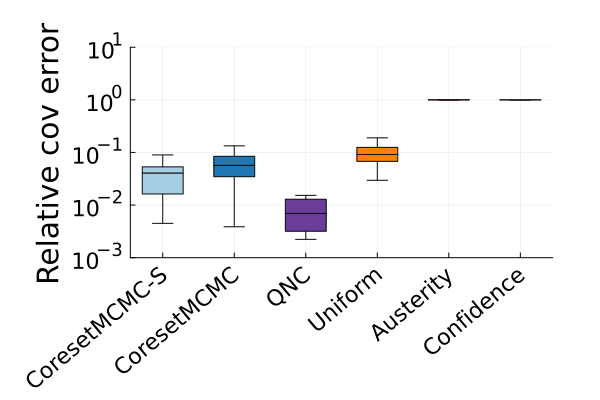}}
    \caption{(h) sparse regression \label{fig:sp_srel_all}}
\end{subfigure}

\caption{Comparison of coreset and subsampling MCMC methods. \cref{fig:lin_mrel_all,fig:log_mrel_all,fig:poi_mrel_all,fig:sp_mrel_all} 
show posterior approximation
quality via the relative mean error, and \cref{fig:lin_srel_all,fig:log_srel_all,fig:poi_srel_all,fig:sp_srel_all}
show posterior approximation quality via the relative covariance error.
The boxplots indicate the median, 25$^\text{th}$, and 75$^\text{th}$ percentiles from 10 runs.}
\label{fig:msrel}
\end{figure*}

\section{\uppercase{Additional results}}\label{sec:additional_results}

In this section we present additional experimental results that show the posterior 
relative mean error ($\|\mu - \hat{\mu}\|_2 / \|\mu\|_2$) and 
relative covariance error ($\|\Sigma - \hat{\Sigma}\|_F / \|\Sigma\|_F$) across all methods.
Similar to the main text, here we use $\hat{\mu}, \hat{\Sigma}$ to denote the mean and covariance estimated 
using draws from each method, and $\mu, \Sigma$ are the same estimated for the full data posterior. Recall that 
the two-moment KL metric that we show in the main text combines both the posterior mean and covariance error 
into a single number. 

\cref{fig:msrel_coreset} shows the relative mean and covariance error between coreset methods for all four models.
Note that for the sparse regression model, this error is computed only on the continuous components.
We see that the trends are mostly similar compared to those of two-moment KLs. 
However, we note that while \texttt{SHF} is able to capture the mean, it approximates the posterior covariances poorly, 
which is due to the limited expressiveness of the variational family. 
All other coreset methods involve using MCMC to sample from the coreset posterior, and so the covariance estimates 
become closer to the true posterior as we increase the coreset size.

We now look at \cref{fig:msrel} to compare the relative mean and covariance error between coreset methods and 
subsampling MCMC methods. We set the coreset size $M=500$ for all coreset methods.
We see that for models with only continuous variables, the lower two-moment KLs from \texttt{SGHMC-CV} that we show in 
the main text largely come from it being able to approximate the posterior mean well.
However, in the Poisson regression example, stochastic optimization 
struggles to identify the mode of the posterior within the time for \texttt{CoresetMCMC} to train and sample.
This causes the posterior approximation quality, especially that of the posterior covariance 
(as shown in \cref{fig:poi_srel_all}), to drop. 

%% file: main.bbl
\begin{thebibliography}{39}
\providecommand{\natexlab}[1]{#1}
\providecommand{\url}[1]{\texttt{#1}}
\expandafter\ifx\csname urlstyle\endcsname\relax
  \providecommand{\doi}[1]{doi: #1}\else
  \providecommand{\doi}{doi: \begingroup \urlstyle{rm}\Url}\fi

\bibitem[Robert and Casella(2004)]{robert1999monte}
Christian Robert and George Casella.
\newblock \emph{{M}onte {C}arlo {S}tatistical {M}ethods}.
\newblock Springer, $2^\text{nd}$ edition, 2004.

\bibitem[Robert and Casella(2011)]{robert2011short}
Christian Robert and George Casella.
\newblock A short history of {M}arkov chain {M}onte {C}arlo: subjective
  recollections from incomplete data.
\newblock \emph{Statistical Science}, 26\penalty0 (1):\penalty0 102--115, 2011.

\bibitem[Gelman et~al.(2013)Gelman, Carlin, Stern, Dunson, Vehtari, and
  Rubin]{gelman2013bayesian}
Andrew Gelman, John Carlin, Hal Stern, David Dunson, Aki Vehtari, and Donald
  Rubin.
\newblock \emph{{B}ayesian {D}ata {A}nalysis}.
\newblock CRC Press, $3^\text{rd}$ edition, 2013.

\bibitem[Banterle et~al.(2019)Banterle, Grazian, Lee, and
  Robert]{banterle2019accelerating}
Marco Banterle, Clara Grazian, Anthony Lee, and Christian Robert.
\newblock Accelerating {M}etropolis-{H}astings algorithms by delayed
  acceptance.
\newblock \emph{Foundations of Data Science}, 1\penalty0 (2):\penalty0
  103--128, 2019.

\bibitem[Quiroz et~al.(2019)Quiroz, Kohn, Villani, and
  Tran]{quiroz2018speeding}
Matias Quiroz, Robert Kohn, Mattias Villani, and Minh-Ngoc Tran.
\newblock Speeding up {MCMC} by efficient data subsampling.
\newblock \emph{Journal of the American Statistical Association}, 114\penalty0
  (526):\penalty0 831--843, 2019.

\bibitem[Maclaurin and Adams(2014)]{maclaurin2014firefly}
Dougal Maclaurin and Ryan Adams.
\newblock Firefly {M}onte {C}arlo: exact {MCMC} with subsets of data.
\newblock In \emph{Uncertainty in Artificial Intelligence}, 2014.

\bibitem[Korattikara et~al.(2014)Korattikara, Chen, and
  Welling]{korattikara2014austerity}
Anoop Korattikara, Yutian Chen, and Max Welling.
\newblock Austerity in {MCMC} land: cutting the {M}etropolis-{H}astings budget.
\newblock In \emph{International Conference on Machine Learning}, 2014.

\bibitem[Bardenet et~al.(2014)Bardenet, Doucet, and
  Holmes]{bardenet2014towards}
R{\'e}mi Bardenet, Arnaud Doucet, and Chris Holmes.
\newblock Towards scaling up {M}arkov chain {M}onte {C}arlo: an adaptive
  subsampling approach.
\newblock In \emph{International Conference on Machine Learning}, 2014.

\bibitem[Welling and Teh(2011)]{welling2011bayesian}
Max Welling and Yee~Whye Teh.
\newblock Bayesian learning via stochastic gradient {L}angevin dynamics.
\newblock In \emph{International Conference on Machine Learning}, 2011.

\bibitem[Chen et~al.(2014)Chen, Fox, and Guestrin]{chen2014stochastic}
Tianqi Chen, Emily Fox, and Carlos Guestrin.
\newblock Stochastic gradient {H}amiltonian {M}onte {C}arlo.
\newblock In \emph{International Conference on Machine Learning}, 2014.

\bibitem[Baker et~al.(2019)Baker, Fearnhead, Fox, and Nemeth]{baker2019sgmcmc}
Jack Baker, Paul Fearnhead, Emily Fox, and Christopher Nemeth.
\newblock \texttt{sgmcmc}: an {R} package for stochastic gradient {M}arkov
  chain {M}onte {C}arlo.
\newblock \emph{Journal of Statistical Software}, 91\penalty0 (3):\penalty0
  1--27, 2019.

\bibitem[Bardenet et~al.(2017)Bardenet, Doucet, and Holmes]{bardenet2017markov}
R{\'e}mi Bardenet, Arnaud Doucet, and Chris Holmes.
\newblock On {M}arkov chain {M}onte {C}arlo methods for tall data.
\newblock \emph{Journal of Machine Learning Research}, 18\penalty0
  (47):\penalty0 1--43, 2017.

\bibitem[Quiroz et~al.(2018)Quiroz, Villani, Kohn, Tran, and
  Dang]{quiroz2018subsampling}
Matias Quiroz, Mattias Villani, Robert Kohn, Minh-Ngoc Tran, and Khue-Dung
  Dang.
\newblock Subsampling {MCMC} - an introduction for the survey statistician.
\newblock \emph{Sankhya A}, 80\penalty0 (1):\penalty0 33--69, 2018.

\bibitem[Nemeth and Fearnhead(2021)]{nemeth2021stochastic}
Christopher Nemeth and Paul Fearnhead.
\newblock Stochastic gradient {M}arkov chain {M}onte {C}arlo.
\newblock \emph{Journal of the American Statistical Association}, 116\penalty0
  (533):\penalty0 433--450, 2021.

\bibitem[Huggins et~al.(2016)Huggins, Campbell, and
  Broderick]{huggins2016coresets}
Jonathan Huggins, Trevor Campbell, and Tamara Broderick.
\newblock Coresets for scalable {B}ayesian logistic regression.
\newblock In \emph{Advances in Neural Information Processing Systems}, 2016.

\bibitem[Campbell and Broderick(2019)]{campbell2019automated}
Trevor Campbell and Tamara Broderick.
\newblock Automated scalable {B}ayesian inference via {H}ilbert coresets.
\newblock \emph{The Journal of Machine Learning Research}, 20\penalty0
  (1):\penalty0 551--588, 2019.

\bibitem[Campbell and Broderick(2018)]{campbell2018bayesian}
Trevor Campbell and Tamara Broderick.
\newblock Bayesian coreset construction via greedy iterative geodesic ascent.
\newblock In \emph{International Conference on Machine Learning}, 2018.

\bibitem[Campbell and Beronov(2019)]{campbell2019sparse}
Trevor Campbell and Boyan Beronov.
\newblock Sparse variational inference: {B}ayesian coresets from scratch.
\newblock In \emph{Advances in Neural Information Processing Systems}, 2019.

\bibitem[Manousakas et~al.(2020)Manousakas, Xu, Mascolo, and
  Campbell]{manousakas2020pseudocoresets}
Dionysis Manousakas, Zuheng Xu, Cecilia Mascolo, and Trevor Campbell.
\newblock {B}ayesian pseudocoresets.
\newblock In \emph{Advances in Neural Information Processing Systems}, 2020.

\bibitem[Naik et~al.(2022)Naik, Rousseau, and Campbell]{naik2022fast}
Cian Naik, Judith Rousseau, and Trevor Campbell.
\newblock Fast {B}ayesian coresets via subsampling and quasi-{N}ewton
  refinement.
\newblock In \emph{Advances in Neural Information Processing Systems}, 2022.

\bibitem[Chen et~al.(2022)Chen, Xu, and Campbell]{chen2022bayesian}
Naitong Chen, Zuheng Xu, and Trevor Campbell.
\newblock Bayesian inference via sparse {H}amiltonian flows.
\newblock In \emph{Advances in Neural Information Processing Systems}, 2022.

\bibitem[Zhang et~al.(2021)Zhang, Khanna, Kyrillidis, and
  Koyejo]{zhang2021revisiting}
Jacky Zhang, Rajiv Khanna, Anastasios Kyrillidis, and Oluwasanmi Koyejo.
\newblock {B}ayesian coresets: revisiting the nonconvex optimization
  perspective.
\newblock In \emph{Artificial Intelligence and Statistics}, 2021.

\bibitem[Benveniste et~al.(1990)Benveniste, M{\'e}tivier, and
  Priouret]{benveniste2012adaptive}
Albert Benveniste, Michel M{\'e}tivier, and Pierre Priouret.
\newblock \emph{Adaptive Algorithms and Stochastic Approximations}.
\newblock Springer, $1^\text{st}$ edition, 1990.

\bibitem[Andrieu and Thoms(2008)]{andrieu2008tutorial}
Christophe Andrieu and Johannes Thoms.
\newblock A tutorial on adaptive {MCMC}.
\newblock \emph{Statistics and Computing}, 18\penalty0 (4):\penalty0 343--373,
  2008.

\bibitem[Robbins and Monro(1951)]{robbins1951stochastic}
Herbert Robbins and Sutton Monro.
\newblock A stochastic approximation method.
\newblock \emph{The Annals of Mathematical Statistics}, 22\penalty0
  (3):\penalty0 400--407, 1951.

\bibitem[Bottou(2004)]{Bottou2004}
L{\'e}on Bottou.
\newblock Stochastic learning.
\newblock In Olivier Bousquet, Ulrike von Luxburg, and Gunnar R{\"a}tsch,
  editors, \emph{Advanced Lectures on Machine Learning: ML Summer Schools
  2003}, chapter~7, pages 146--168. Springer Berlin Heidelberg, 2004.

\bibitem[Kingma and Ba(2014)]{kingma2014adam}
Diederik Kingma and Jimmy Ba.
\newblock Adam: a method for stochastic optimization.
\newblock In \emph{International Conference on Learning Representations}, 2014.

\bibitem[Duchi et~al.(2011)Duchi, Hazan, and Singer]{duchi2011adaptive}
John Duchi, Elad Hazan, and Yoram Singer.
\newblock Adaptive subgradient methods for online learning and stochastic
  optimization.
\newblock \emph{Journal of Machine Learning Research}, 12\penalty0
  (61):\penalty0 2121--2159, 2011.

\bibitem[Neal(2003)]{neal2003slice}
Radford Neal.
\newblock Slice sampling.
\newblock \emph{The Annals of Statistics}, 31\penalty0 (3):\penalty0 705--767,
  2003.

\bibitem[Neal(2011)]{neal2011mcmc}
Radford Neal.
\newblock {MCMC} using {H}amiltonian dynamics.
\newblock In Steve Brooks, Andrew Gelman, Galin Jones, and Xiao-Li Meng,
  editors, \emph{Handbook of {M}arkov chain {M}onte {C}arlo}, chapter~5. CRC
  Press, 2011.

\bibitem[Duane et~al.(1987)Duane, Kennedy, Pendleton, and
  Roweth]{duane1987hybrid}
Simon Duane, Anthony Kennedy, Brian Pendleton, and Duncan Roweth.
\newblock Hybrid {M}onte {C}arlo.
\newblock \emph{Physics Letters B}, 195\penalty0 (2):\penalty0 216--222, 1987.

\bibitem[B{\'e}lisle et~al.(1993)B{\'e}lisle, Romeijn, and
  Smith]{belisle1993hit}
Claude B{\'e}lisle, Edwin Romeijn, and Robert Smith.
\newblock Hit-and-run algorithms for generating multivariate distributions.
\newblock \emph{Mathematics of Operations Research}, 18\penalty0 (2):\penalty0
  255--266, 1993.

\bibitem[Rakhlin et~al.(2012)Rakhlin, Shamir, and Sridharan]{rakhlin2011making}
Alexander Rakhlin, Ohad Shamir, and Karthik Sridharan.
\newblock Making stochastic gradient descent optimal for strongly convex
  problems.
\newblock In \emph{International Conference on Machine Learning}, 2012.

\bibitem[Hazan and Kale(2014)]{hazan2014beyond}
Elad Hazan and Satyen Kale.
\newblock Beyond the regret minimization barrier: optimal algorithms for
  stochastic strongly-convex optimization.
\newblock \emph{The Journal of Machine Learning Research}, 15\penalty0
  (1):\penalty0 2489--2512, 2014.

\bibitem[Gower et~al.(2020)Gower, Schmidt, Bach, and
  Richt{\'a}rik]{gower2020variance}
Robert Gower, Mark Schmidt, Francis Bach, and Peter Richt{\'a}rik.
\newblock Variance-reduced methods for machine learning.
\newblock \emph{Proceedings of the IEEE}, 108\penalty0 (11):\penalty0
  1968--1983, 2020.

\bibitem[Carpenter et~al.(2017)Carpenter, Gelman, Hoffman, Lee, Goodrich,
  Betancourt, Brubaker, Guo, Li, and Riddell]{carpenter2017stan}
Bob Carpenter, Andrew Gelman, Matthew Hoffman, Daniel Lee, Ben Goodrich,
  Michael Betancourt, Marcus Brubaker, Jiqiang Guo, Peter Li, and Allen
  Riddell.
\newblock Stan: a probabilistic programming language.
\newblock \emph{Journal of Statistical Software}, 76\penalty0 (1):\penalty0
  1–--32, 2017.

\bibitem[George and McCulloch(1993)]{george1993variable}
Edward George and Robert McCulloch.
\newblock Variable selection via {G}ibbs sampling.
\newblock \emph{Journal of the American Statistical Association}, 88\penalty0
  (423):\penalty0 881--889, 1993.

\bibitem[Vehtari et~al.(2021)Vehtari, Gelman, Simpson, Carpenter, and
  B{\"u}rkner]{vehtari2021rank}
Aki Vehtari, Andrew Gelman, Daniel Simpson, Bob Carpenter, and Paul-Christian
  B{\"u}rkner.
\newblock Rank-normalization, folding, and localization: an improved
  $\widehat{R}$ for assessing convergence of {MCMC} (with discussion).
\newblock \emph{Bayesian Analysis}, 16\penalty0 (2):\penalty0 667--718, 2021.

\bibitem[Wolfe(1969)]{wolfe1969convergence}
Philip Wolfe.
\newblock Convergence conditions for ascent methods.
\newblock \emph{SIAM Review}, 11\penalty0 (2):\penalty0 226--235, 1969.

\end{thebibliography}
